\pdfoutput=1

\documentclass[epj]{svjour}
\usepackage{graphics}
\usepackage{listings}
\usepackage{subfig}
\usepackage{graphicx}
\usepackage{amsmath}
\usepackage{epstopdf}
\usepackage{epsfig}
\usepackage{xcolor}
\usepackage{hyperref} 
\usepackage{units} 
\begin{document}
\title{Measurement of the Muon Production Depths at the Pierre Auger Observatory}
\author{Laura Collica\inst{1} for the Pierre Auger Collaboration\inst{2}
}                     
\offprints{}          
\institute{INFN Torino, Via Pietro Giuria 1, Italy \and Observatorio Pierre Auger, Av.\ San Mart\'in Norte 304, 5613 Malarg\"ue, Argentina \\
Full author list: \href{http://www.auger.org/archive/authors_2015_06.html}{\rm http://www.auger.org/archive/authors\_2016\_05.html}}

\date{Date: 6 September 2016 \\ Published in EPJ as DOI 10.1140/epjp/i2016-16301-6}
%

\abstract{
The muon content of extensive air showers is an observable sensitive to the primary composition and to the hadronic interaction properties. The Pierre Auger Observatory uses water-Cherenkov detectors to measure particle densities at the ground and therefore is sensitive to the muon content of air showers. We present here a method which allows us to estimate the muon production depths by exploiting the measurement of the muon arrival times at the ground recorded with the Surface Detector of the Pierre Auger Observatory. The analysis is performed in a large range of zenith angles, thanks to the capability of estimating and subtracting the electromagnetic component, and for energies between $10^{19.2}$ and $10^{20}$ eV.
} 
\authorrunning{L. Collica for the Pierre Auger Collaboration}
\titlerunning{Measurement of the Muon Production Depths at Auger}
\maketitle
\section{Introduction}
\label{intro}
The spectrum and arrival directions of Ultra High Energy Cosmic Rays (UHECRs) above $10^{18}$ eV have been recently measured with unprecedented precision \cite{bib:valino,bib:arr}. The flux of cosmic rays at these energies is very low (less than 100 particles $\mathrm{km^{-2} yr^{-1}}$) and their origin is still not well understood. 
Establishing the cosmic-ray composition at the highest energies is of fundamental importance from the astrophysical point of view, since it could discriminate between different scenarios of origin and propagation of cosmic rays. Moreover, mass composition studies are of utmost importance for particle physics. As a matter of fact, knowing the composition helps in exploring the hadronic interactions at ultra-high energies, inaccessible to present accelerator experiments. \\
UHECRs properties cannot be determined from direct detection, due to their low flux, but must be inferred from the measurements of the secondary particles that the cosmic-ray primary produces in the atmosphere. These particles cascades are called Extensive Air Showers (EAS) and can be studied at the ground by deploying detectors covering large areas.  \\
Composition studies on a shower to shower basis are challenging because of the intrinsic shower-to-shower fluctuations which characterise shower properties. These fluctuations come from the random nature of the interaction processes, in particular the height of the first interaction. However, showers originating from different primaries can be distinguished, at least statistically, given their different cross sections with air nuclei and distinct hadronic multiparticle production properties.
Masses may be inferred from comparisons of the measured observables with predictions for these same observables from Monte Carlo simulations. These simulations rely on hadronic interaction models, which extrapolate interaction details from measurements in the accelerators domain to much higher energies and to different kinematic regions. Therefore, comparisons with simulations constitute the most prominent source of systematic uncertainties. \\
Information about the composition of the primary cosmic rays has been obtained using the Fluorescence Detector (FD) of the Pierre Auger Observatory \cite{bib:auger}. The FD allows the measurement of the depth at which the electromagnetic component of the air shower reaches its maximum number of particles, $X_\mathrm{max}$ \cite{bib:aab}.
This observable is sensitive to the nature of the primary particles, as well as the standard deviation of its distribution, $\sigma(X_\mathrm{max})$. 
The interpretation of these measurements is hampered by uncertainties in hadronic interaction models. Besides, the number of events detected with the FD at high energy is low, due to the small FD duty cycle (about 15\%); the stringent cuts imposed to avoid a biased data sample in the analysis, such a field of view cut, further reduce the available statistics. \\
To gain additional information about mass composition and investigate the validity of the current hadronic interaction models, independent measurements with larger statistics are needed, together with a different set of systematic uncertainties and the possibility of reaching higher energies.  \\
The Pierre Auger Collaboration has developed different methods to infer the composition of UHECRs through the measurements performed with the Surface Detector (SD), which has 100\% duty cycle. Among them, the study of the atmospheric depth at which the muon production rate reaches a maximum in air showers exploits the fact that the muon production depth is one of the most sensitive observables to the primary mass  \cite{bib:MPD}. In addition, muons are sensitive to hadronic interactions since they come from the decay of pions and kaons, which form the hadronic core, and suffer small energy losses and angular deflections on their way to the ground.\\
In this paper a method for the reconstruction of the muon production depth for zenith angles between $45^{\circ}$ and $65^{\circ}$ and energies greater than $10^{19.2}$ eV is presented. The method is based on the model of muon time distributions discussed in \cite{bib:MPD} but it exploits a new kinematic delay parametrisation, tuned on post-LHC models, and a different technique to estimate the electromagnetic component. The latter allows one to reconstruct the muon arrival times at the ground closer to the shower core and for lower zenith angles, thus improving the muon sampling and the range of applicability of the analysis. \\
The paper is organised as follows. Sect.~\ref{sec:auger} briefly describes the Pierre Auger Observatory. In Sect.~\ref{sec:model} the muon time distributions model is described in details. Sect.~\ref{sec:featMPD} gives an overview of the properties of the muon production depth distribution while Sect.~\ref{sec:method} discusses all the steps through which the muon production depth is reconstructed. Finally, in Sect.~\ref{sec:sum} a brief summary is given together with some comments on the possible application of the method.

\section{The Pierre Auger Observatory}
\label{sec:auger}

The Pierre Auger Observatory \cite{bib:auger} is the largest operating cosmic-ray observatory ever built. The main goal of the observatory is to address the unsolved questions about UHECR physics. It provides a sample of more than ten years of data, continuously recorded since January 2004. 
The Observatory is located in Argentina near the town of Malarg\"ue (1420 m a.s.l., 870 $\mathrm{g/cm^{2}}$), in the Pampa Amarilla region, and is composed by two types of detectors, making it an hybrid experiment. \\
The Surface Detector (SD) \cite{bib:sd} consists of 1660 cylindrical water-Cherenkov detectors arranged on a triangular grid, with 1500 m spacing, that covers an area of about 3000 $\mathrm{km^2}$. Each station is filled with 12 tonnes of purified water; three 9-inch photomultipliers (PMTs) detect the Cherenkov light produced in water by the secondary particles of the air showers. The signals from the three PMTs are obtained using Flash Analog to Digital Converters (FADC) that process them at 40 MHz sampling rate.\\
The detector is sensitive to muons and electromagnetic particles and is calibrated in units of the signal produced by a muon traversing the water vertically, known as a \textit{Vertical Equivalent Muon} (VEM). An SD event is formed when at least three non-aligned stations selected by the local station trigger are in spatial and temporal coincidence. \\
The Fluorescence Detector (FD) \cite{bib:fd,bib:heat} is composed of 27 optical telescopes placed at four different locations. On clear moonless nights the FD observes the atmosphere above the SD, allowing for an hybrid detection of Extensive Air Showers (EAS). In particular the FD measures the longitudinal development of showers by detecting the fluorescence and Cherenkov light produced in the atmosphere by charged particles along the shower trajectory. It thus provides a calorimetric measurement of the primary particle energy.
An event detected by at least one FD telescope and one SD station is called \textit{hybrid}. The combination of the timing information from the FD and the SD provides an accurate determination of the geometry of the air showers. 
 A detailed description of detectors and of reconstruction procedures can be found in \cite{bib:auger}. \\
For energies greater than 3 EeV, the SD detection efficiency is 100\%. The arrival directions are obtained
from the times at which the shower front passes through the triggered detectors, this time being measured using GPS information with an accuracy of 10 ns. The angular resolution, defined as the angular radius around the true cosmic-ray direction that would contain 68\% of the reconstructed shower directions, is $0.8^{\circ}$ for energies above 3 EeV \cite{bib:bon2008}.\\
A direct energy calibration is provided by hybrid events and the energy resolution above 10 EeV is about 12\%. The absolute energy scale, determined by the FD, has a systematic uncertainty of about 14\%.\\
In the context of primary mass studies, hybrid events have been used to provide a direct measurement of $X_{\mathrm{max}}$.  However, the bulk of events recorded by the observatory has information only from the surface array, making SD observables crucial for composition analyses at the highest energies.

\section{The model of the muon time distributions} \label{sec:model}

Muons mainly come from the decay of pions and kaons which are produced with a characteristic transverse momentum distribution inside a narrow cylinder around the shower axis (for a 50 GeV pion, the muon production angle is $\sim 2^{\circ}$).  \\
Since the muon radiation length is much larger than the whole atmospheric depth, even in the case of inclined events, bremsstrahlung and pair production are improbable, and multiple scattering effects are negligible. It could thus be assumed that muons travel following straight lines, from the point where they are produced. \\
The time structure of the muon component reaching the ground can be exploited to obtain the distribution of muon production distances along the shower axis, which provides information about the longitudinal development of the hadronic component of the EAS. This information is complementary to that obtained from the electromagnetic component through the detection of the atmospheric fluorescence light. \\
At the ground, the time structure of muons is the convolution of the production spectra, the energy loss, and the decay probability during propagation. A phenomenological model for muon time distributions in EAS was developed in \cite{bib:mod-old,bib:cazon-mod} and its main characteristics are summarised below.\\
\begin{figure}
\begin{minipage}[b]{.40\textwidth}
\centering
\includegraphics [width=1.15\textwidth]{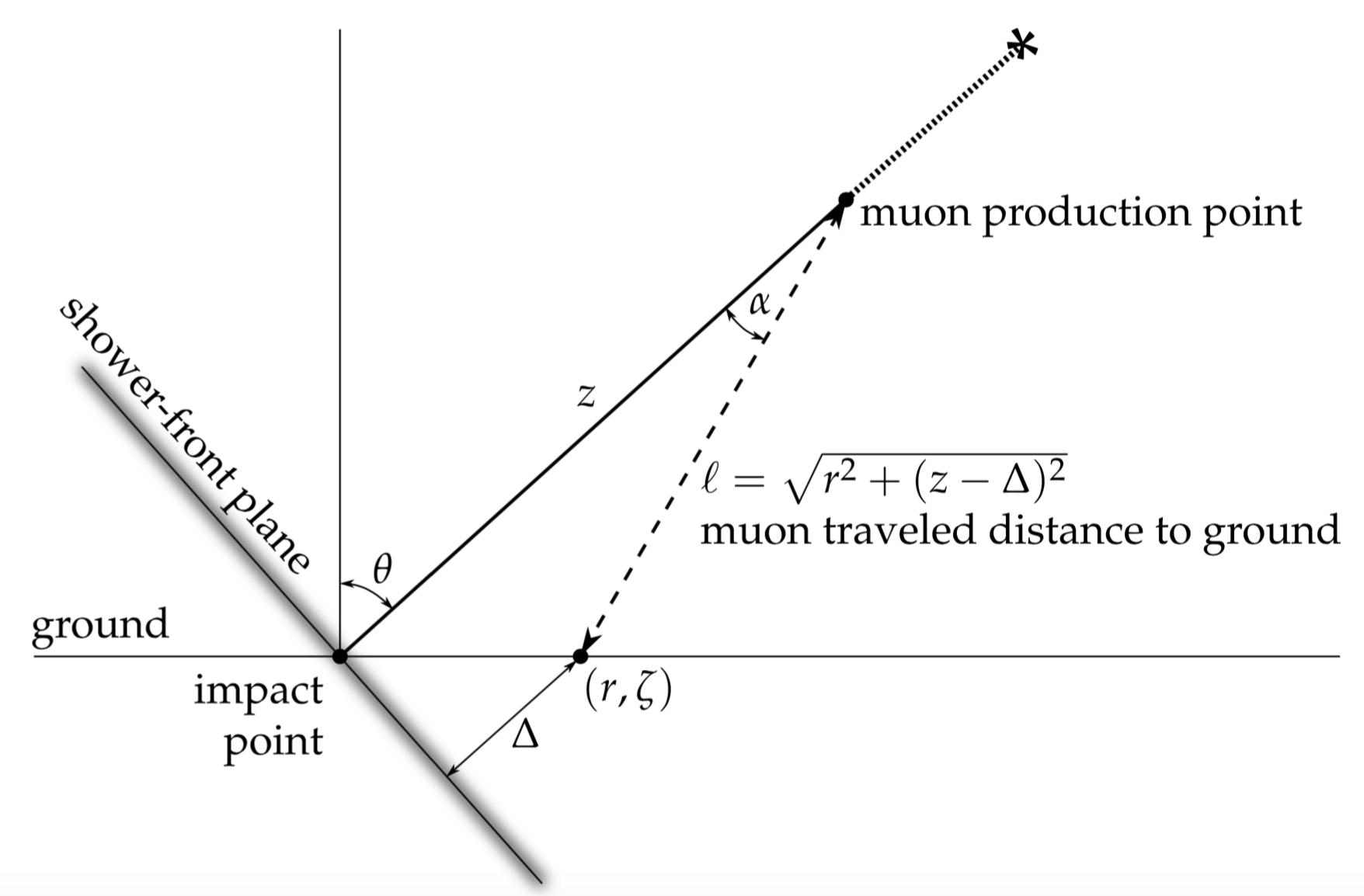} 
\end{minipage}%
\hspace{20mm}%
\begin{minipage}[b]{.40\textwidth}
\includegraphics [width=1.15\textwidth]{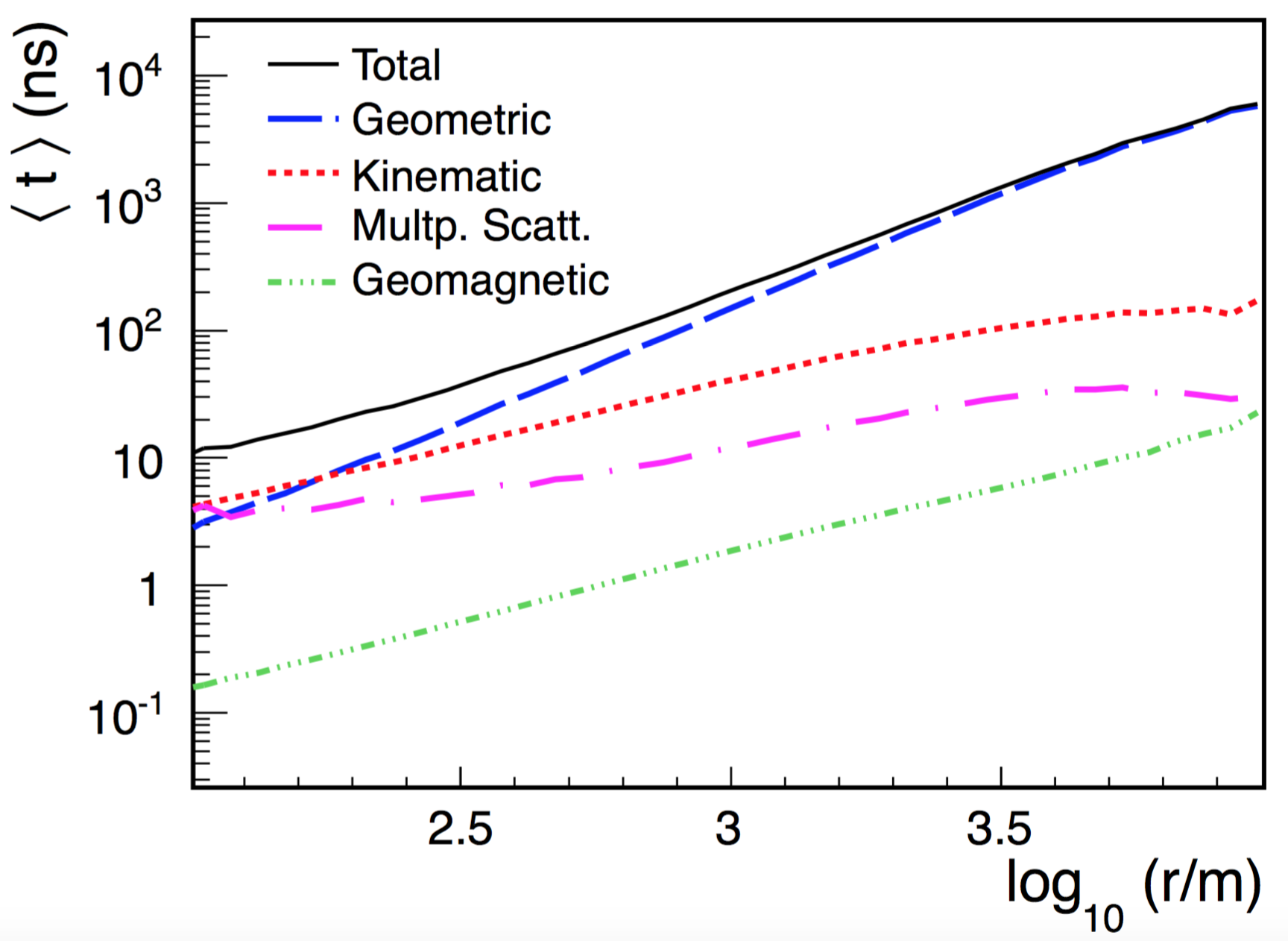}
\end{minipage}
\caption{\small Left) The geometry used to obtain the muon traveled distance and the time delay. Right) Average delays as a function of distance from the core for $10^{19}$ eV proton showers at $60^{\circ}$. The total delay and its contributions are shown \cite{bib:cazon-mod}.}
\label{geomdelay}
\end{figure}
The geometry used to obtained the muon production point is outlined in Fig.~\ref{geomdelay} (left). Muons are produced at the position $z$ along the shower axis and, after traveling a distance $l$, they reach the ground at the point defined by $(r,\zeta)$.
The variables $r$ and $\zeta$ are measured in the shower reference frame and represent the distance and the azimuthal position of the point at the ground, respectively.
The muon production point along the shower axis $z$ can be written as 
\begin{equation}
z \simeq \frac{1}{2} \Biggl ( \frac{r^{2}}{c t_{g}} -c t_{g} \Biggr )+ \Delta - \langle z_{\pi} \rangle
\label{eq:z-for}
\end{equation}
where ${t_{g}\simeq t - \langle t_{\epsilon} \rangle}$ is the \textit{geometric delay}, due to the deviation of muon trajectories with respect to the shower axis, $t$ is the measured muon delay with respect to the shower front, $t_{\epsilon}$ is the \textit{kinematic delay} due to the sub-luminal muon velocities, $\Delta = r \tan \theta \cos \zeta$ is the distance from the impact point at the ground to the shower plane, and $\langle z_{\pi} \rangle$ takes into account the decay length of the parent pion. Indeed, muons are not produced on the shower axis, but collinear with the trajectory followed by the parent pions. The muon path starts deeper in the atmosphere by an amount which is simply the decay length of the pion: $z_{\pi}=c\tau_{\pi}E_{\pi}/(m_{\pi}c^2)\cos \alpha$. The pion energy dependence of this correction has been taken from \cite{bib:mod-old2}. The angle $\alpha$ is estimated for each muon by exploiting the geometry shown in Fig.~\ref{geomdelay} (left). The distance $\langle z_{\pi} \rangle$ introduces an average time delay of $\sim$ 3 ns (this correction amounts to $\sim$1\% of the total delay).\\
The measured muon delay is not only due to the geometric delay. Indeed, muons do not travel at the speed of light and they lose energy on their way to the ground, mainly because of inelastic collisions with atomic electrons in the air. This contribution to the total delay, called kinematic delay, can be estimated from the muon energy \cite{bib:mod-old2}. Since this is not a measurable observable in SD stations, a parametrisation of the kinematic delay must be studied (see Sect. \ref{sec:param}). \\
Two additional sources of delay are given by the deflection of muons due to their elastic scattering off nuclei and by the delaying of muons due to the geomagnetic field, which affects their trajectory. \\
The evolution of the total delay and its sources as a function of distance from the core is shown in Fig.~\ref{geomdelay} (right): the geometric delay dominates far from the shower core, while the kinematic one is larger near the core. Indeed, in this region the spread in muon energy is larger and the mean time delay is dominated by low-energy muons. At distances greater than about 1200 m the kinematic delay is less than 20\% of the geometric delay, while the other contributions are of the order of a few percent. This is valid for events with different zenith angles.\\
Eq. \ref{eq:z-for} provides the muon production point $z$ for each muon delay $t$ measured at the ground. The production depth $X^{\mu}$, i.e. the total amount of traversed matter in $\mathrm{g/cm^2}$, is easily related to the production distance using
\begin{equation}
X^{\mu}(z) =  \int_{z}^{\infty} \rho(z') \mathrm{d}z'
\label{eq:int}
\end{equation}
where $\rho$ is the atmosphere density. \\
The set of production depths forms the Muon Production Depth (MPD) distribution, which contains information about the longitudinal development of the hadronic cascade. In particular the MPD maximum, $X_{\mathrm{max}}^{\mu}$, is sensitive to the primary mass, as the maximum of the electromagnetic longitudinal distribution measured with FD detectors.

\section{Features of the MPD distribution} \label{sec:featMPD}

\begin{figure}
\begin{minipage}[b]{.40\textwidth}
\centering
\includegraphics [width=1.2\textwidth]{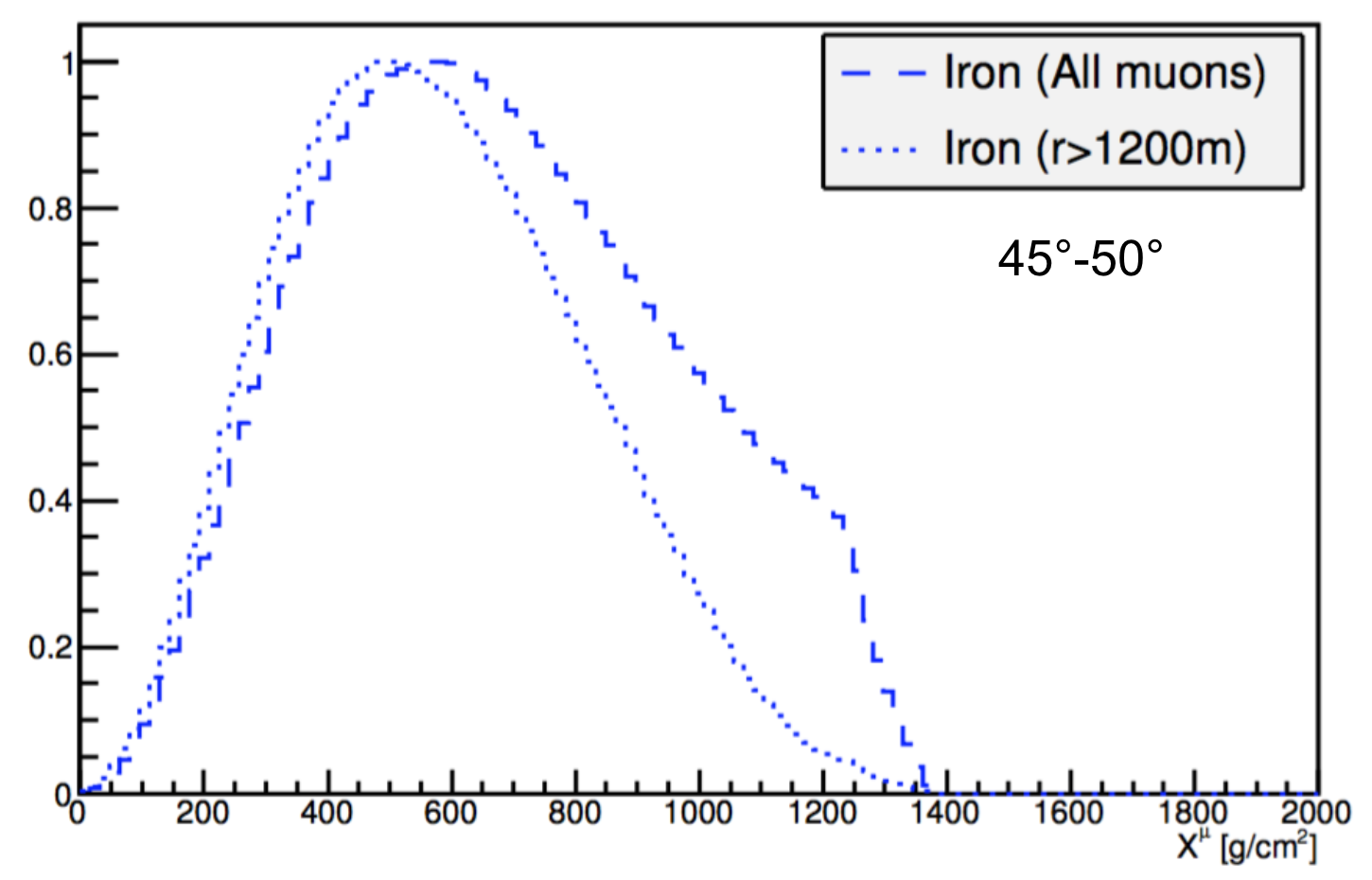} 
\end{minipage}%
\hspace{20mm}%
\begin{minipage}[b]{.40\textwidth}
\includegraphics [width=1.2\textwidth]{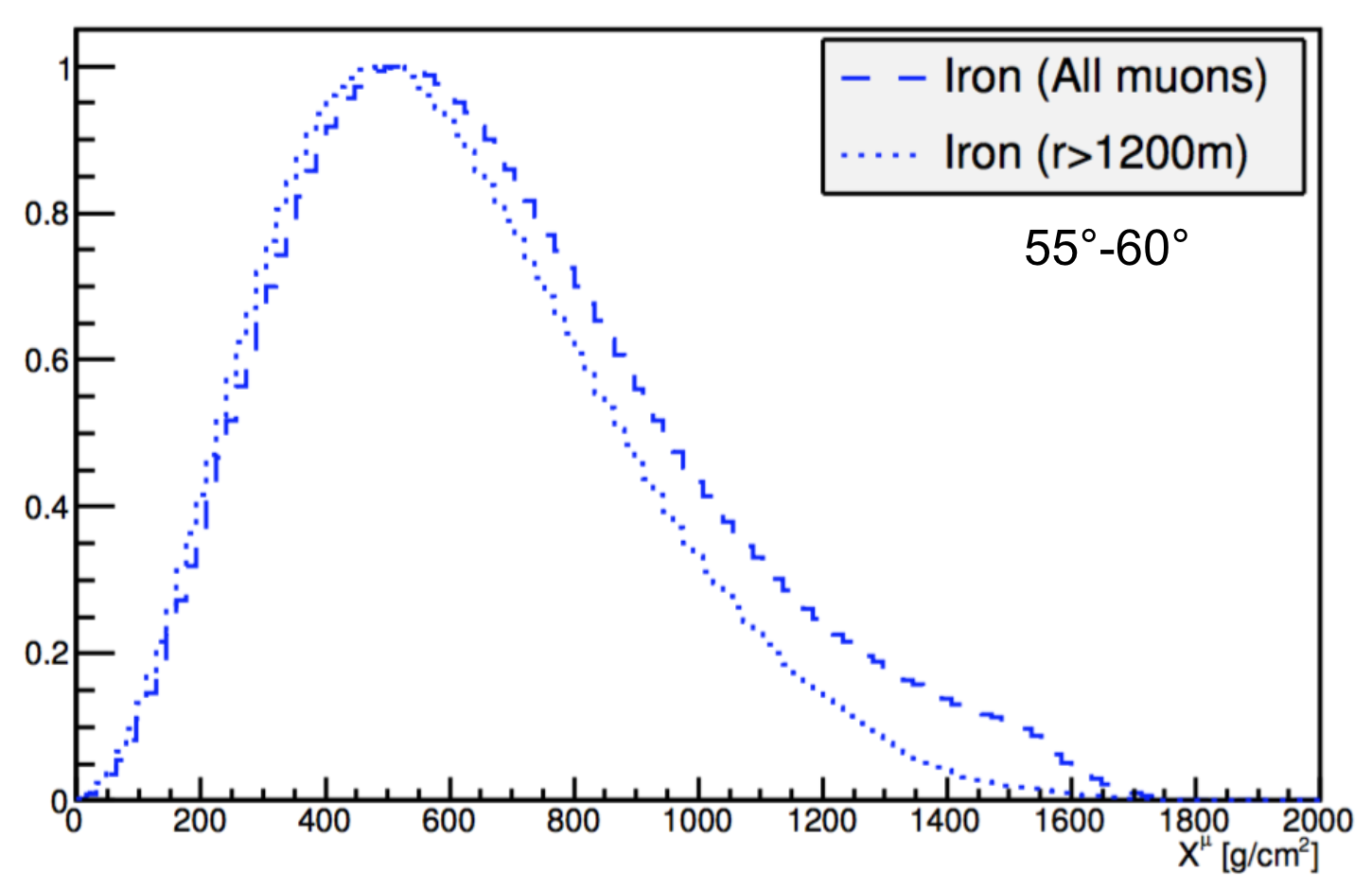}
\end{minipage}
\caption{\small The average MPD distribution without the distance cut (broken line) and with the distance cut $\mathrm{r_{cut}}>1200$ m (dotted line) are shown. Two zenith-angle bins are considered here: $\theta=45^{\circ}-50^{\circ}$ (left) and $\theta=55^{\circ}-60^{\circ}$ (right). The distributions were obtained simulating iron-induced showers  at $E=10^{19.5}$ eV and adopting the model QGSJetII-04 for the hadronic interactions.}
\label{mpd_cut_dist}
\end{figure}

The \textit{true} MPD distribution, i.e. the one relative to all muons produced in air showers, cannot be reconstructed at ground level since a fraction of muons will decay before reaching the ground.
From now on, we will always refer to the \textit{apparent} distribution, which is the one formed by all muons surviving at the atmospheric depth of the experiment. \\
To define the explorable ranges in energy, zenith angle, and distances from the core for the MPD reconstruction, the MPD properties have been investigated exhaustively, using simulations for proton and iron showers done with the code CORSIKA \cite{bib:corsika} and based on the most recent post-LHC hadronic interaction models, EPOS-LHC \cite{bib:epos} and QGSJetII-04 \cite{bib:qgs}. These models do not assume new physical effects in hadronic interactions and are based on cross-sections extrapolated from LHC data. \\
As already discussed, the muon production in EAS is mostly due to charged-pion decay. The critical energy of $\pi^{\pm}$ depends on the atmospheric density: the decay probability of $\pi^{\pm}$ is greater than the interaction one when air density is low.
The MPD distribution thus depends on the zenith angle of the shower: inclined events develop at higher depths than vertical ones since the average density seen by the former is smaller.
At high zenith angles the MPD distribution is well defined while at low angles it undergoes an abrupt truncation since the shower arrives at the ground before reaching the muon maximum. This is especially true for protons, since their showers develop deeper in the atmosphere with respect to iron ones. As a consequence, the analysis is focused on events with zenith angle greater than $45^{\circ}$, for which the distribution maximum is well defined on an event-by-event basis and for both primaries. \\
A limited range in core distance must be considered to keep the contribution of the kinematic delay low, being the latter parametrised. A cut at $r=1200$ m ensures $t_{\epsilon}<0.2 \ t_{g}$. In addition, the trigger request of a minimum of 3 VEM per station sets an upper limit on the considered core distances at $r=4000$ m.
The cut in distance affects the MPD distribution since all muons that arrive close to the core are cut off. The effect of the cut on the MPD distribution is greater for lower zenith angles as shown in Fig.~\ref{mpd_cut_dist}. In particular the distance cut suppresses the tail of the distribution, since the latter is mostly populated by low energy muons produced close to the ground which arrive near the core. \\
With the aim of obtaining useful physical information from the MPD distribution, for each shower we perform a fit of the muon longitudinal development profile with the Universal Shower Profile (USP) function, which describes well its non-symmetric shape \cite{bib:usp}.
This function has three parameters, all related to the physics of the shower: the maximum of the profile $X_{\mathrm{max}}^{\mu}$, the profile width $L$, and a parameter related to the distribution asymmetry $R$, which quantifies the deformation of the profile with respect to a Gaussian distribution.\\
The USP function is defined as
\begin{equation}
\frac{\mathrm{d}N}{\mathrm{d}X} = \left( 1+ \frac{R}{L} (X-X_{\mathrm{max}}^{\mu}) \right)^{R^{-2}} \exp^{-\frac{X-X_{\mathrm{max}}^{\mu}}{LR}},
\label{eq:usp}
\end{equation}
where $N$ refers to the number of produced muons and $X$ is the depth expressed in $\mathrm{g/cm^2}$ \cite{bib:usp2}.
Of the three parameters, $X_{\mathrm{max}}^{\mu}$ accounts for the point along the shower axis where the production of muons reaches its maximum as the shower develops through the atmosphere. This parameter will be our main physical observable for composition and hadronic interactions studies. \\
The best set of parameters that describes a given longitudinal muon profile (either at generation or reconstruction level) is obtained through a log-likelihood minimisation of the USP function. When working with the MPD distributions at generation level, i.e. using the muon production points directly obtained from the simulation code CORSIKA, we fit the profile leaving all the parameters free. \\
In reconstructed events, the number of muons which are used to build the MPD distribution is not very large. For low zenith angles, after cuts, typically about 50 muons (${\sim}$ 10 SD stations) contribute at an energy of $10^{19.5}$ eV. At high zenith angles, the number is higher, about 60 muons for the same energy. \\
Two reasons are at the source of this shortage: on one hand, the detectors are separated by 1500 m and have a finite collecting surface; on the other hand we select stations far from the core (and therefore with small signals) to ensure an accurate determination of $X_{\mathrm{max}}^{\mu}$. This muon undersampling does not yield reliable estimates when all three parameters of the USP function are left free in the minimisation. This is particularly true for proton-induced showers, which suffer most the undersampling at higher production depths (low $X$).
The best fit is thus obtained by setting the starting value of the parameter $L$ and by fixing the asymmetry parameter $R$. 
According to simulations, the parameter $R$, as $L$, depends significantly on the zenith angle and it is not possible to fix it to a single value in the whole zenith angle range. The preferred $R$ value also depends on the nature of the primary particle. As a consequence, we studied the dependence of both $L$ and $R$ on a mixed sample of proton/iron-induced showers simulated with both QGSJetII-04 and EPOS-LHC. In the fitting procedure, the initial value for $L$ is assigned while $R$ is fixed event-to-event, on the basis of their angular dependence. \\
\begin{figure}
\begin{minipage}[b]{.40\textwidth}
\centering
\includegraphics [width=1.25\textwidth]{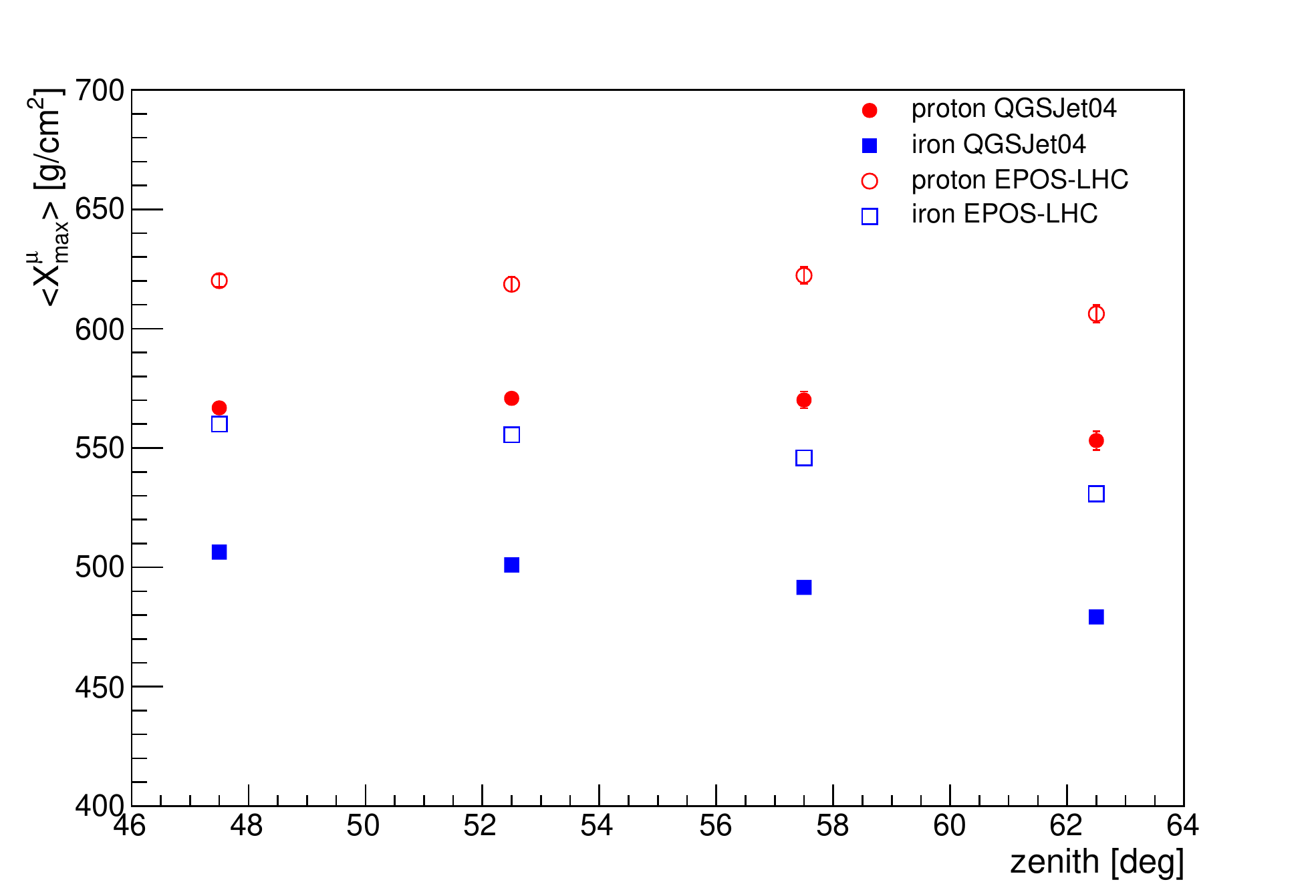} 
\end{minipage}%
\hspace{20mm}%
\begin{minipage}[b]{.40\textwidth}
\includegraphics [width=1.25\textwidth]{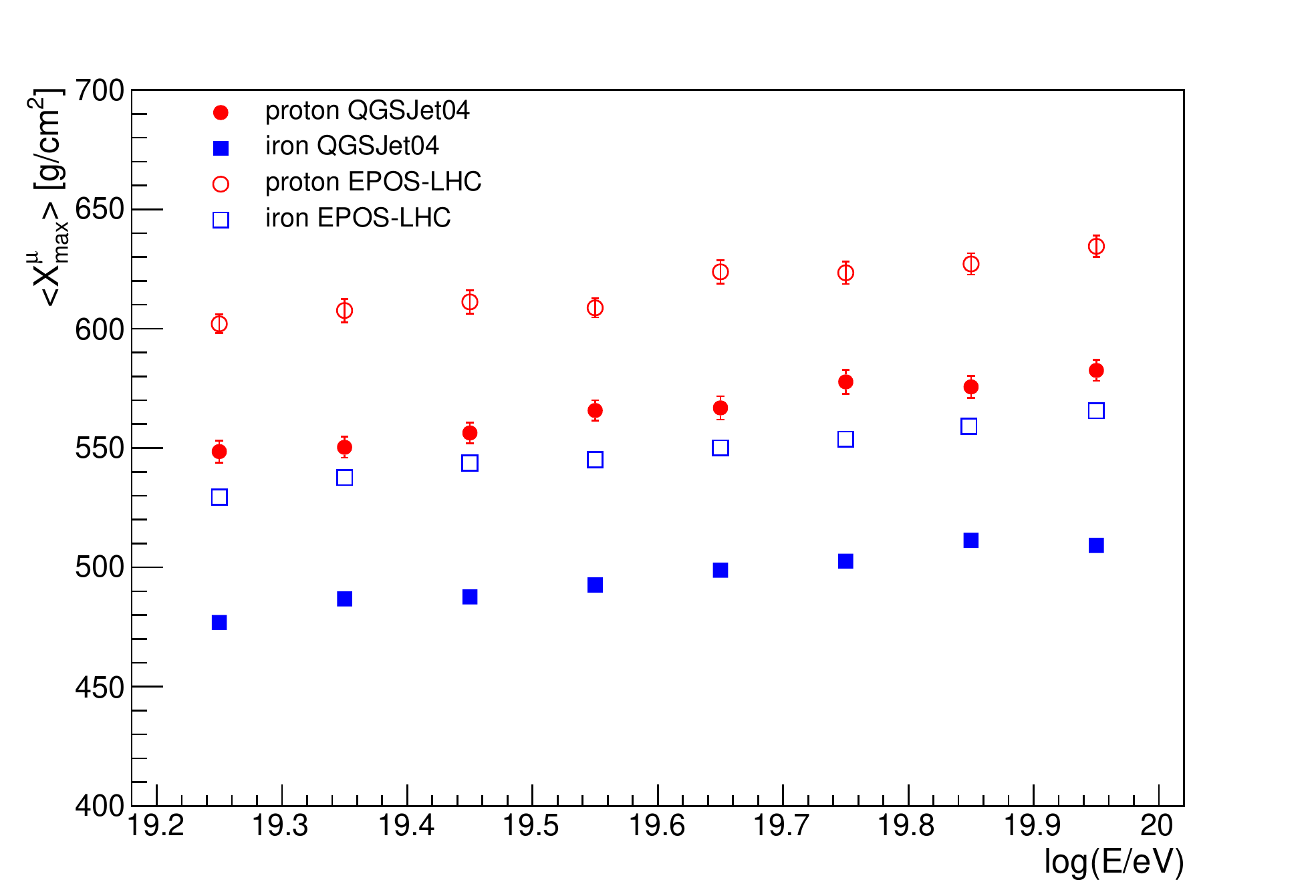}
\end{minipage}
\caption{\small Left) $\langle X_{\mathrm{max}}^{\mu}\rangle$ is shown as a function of the zenith angle. Right) $\langle X_{\mathrm{max}}^{\mu} \rangle$ is shown as a function of the logarithm of energy. Results are shown for proton (red) and iron (blue) showers simulated with the two post-LHC models.}
\label{Xmax_zen_energy}
\end{figure}
The MPD distribution fit is performed in an interval of atmospheric depths ranging from 0 to 1200 $\mathrm{g/cm^2}$, which contains the entire range of possible values of $X_{\mathrm{max}}^{\mu}$ (the largest event has an energy of 96 EeV and $X_{\mathrm{max}}^{\mu} \simeq 1000 \ \mathrm{g/cm^2}$).
In Fig.~\ref{Xmax_zen_energy} (left) the $\langle X_{\mathrm{max}}^{\mu} \rangle$ is shown as a function of the zenith angle, for both masses and hadronic interaction models. It is smaller for an iron than for a proton shower and it slightly decreases with the zenith angle, especially in the case of iron showers.  \\
Taking into account this angular dependence, each $X_{\mathrm{max}}^{\mu}$ can be referred to $\langle \theta \rangle =55^{\circ}$ and its dependence on the primary energy is shown in Fig.~\ref{Xmax_zen_energy} (right). The muon maximum increases with the logarithm of the energy, as its electromagnetic counterpart \cite{bib:heitler,bib:matthews}.
The evolution of  $\langle X_{\mathrm{max}}^{\mu} \rangle$ with the energy is called the \textit{muonic elongation rate}. \\
It is important to note that the two hadronic interaction models predict a similar elongation rate but they show a considerable difference in the absolute value of  $\langle X_{\mathrm{max}}^{\mu} \rangle$. Indeed this difference is almost comparable to the difference in $\langle X_{\mathrm{max}}^{\mu} \rangle$ for proton-induced and iron-induced showers, which is about 70 $\mathrm{g/cm^2}$.
The MPD analysis is thus an optimal tool to put constraints on hadronic interaction models. \\
\begin{figure}
\begin{minipage}[b]{.40\textwidth}
\centering
\includegraphics [width=1.25\textwidth]{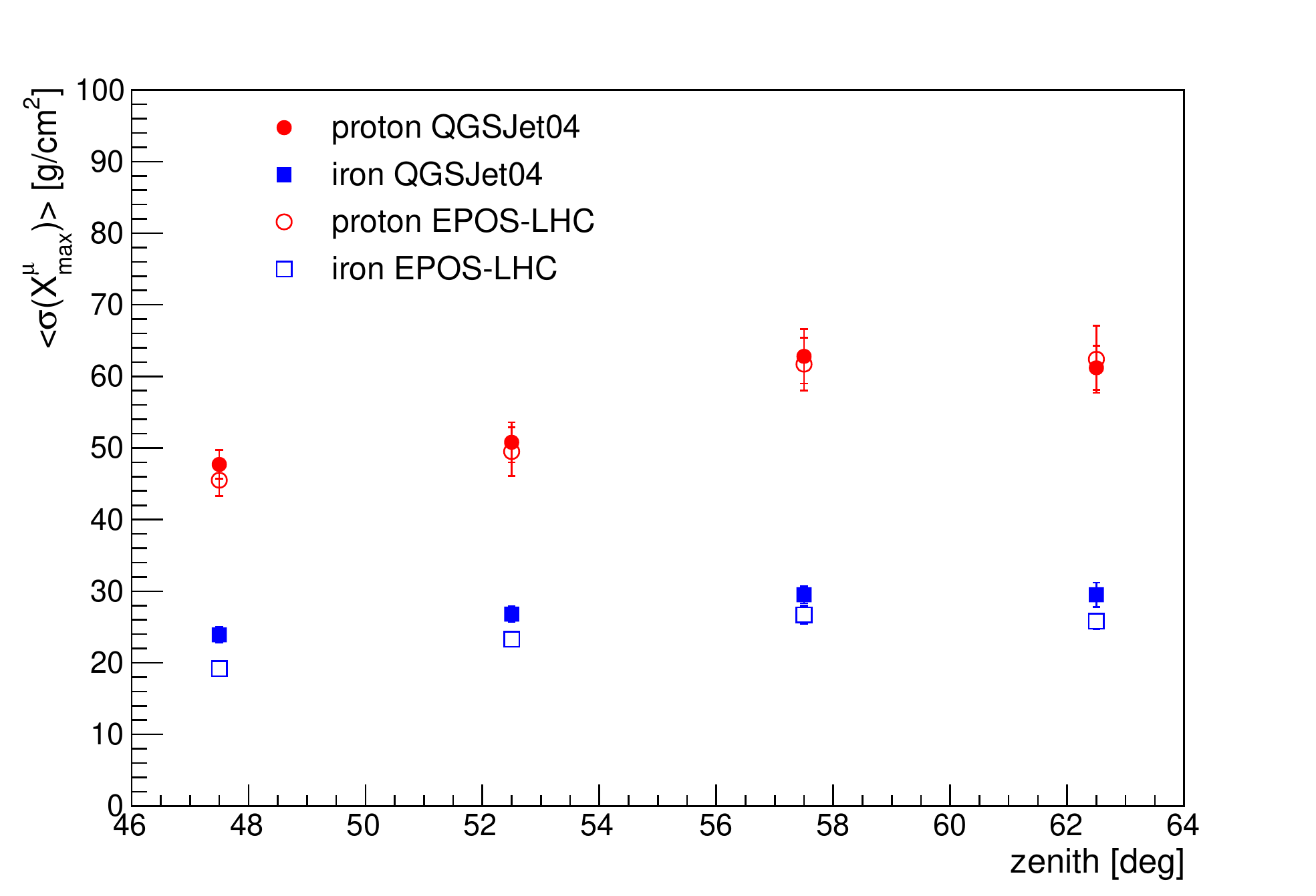} 
\end{minipage}%
\hspace{20mm}%
\begin{minipage}[b]{.40\textwidth}
\includegraphics [width=1.25\textwidth]{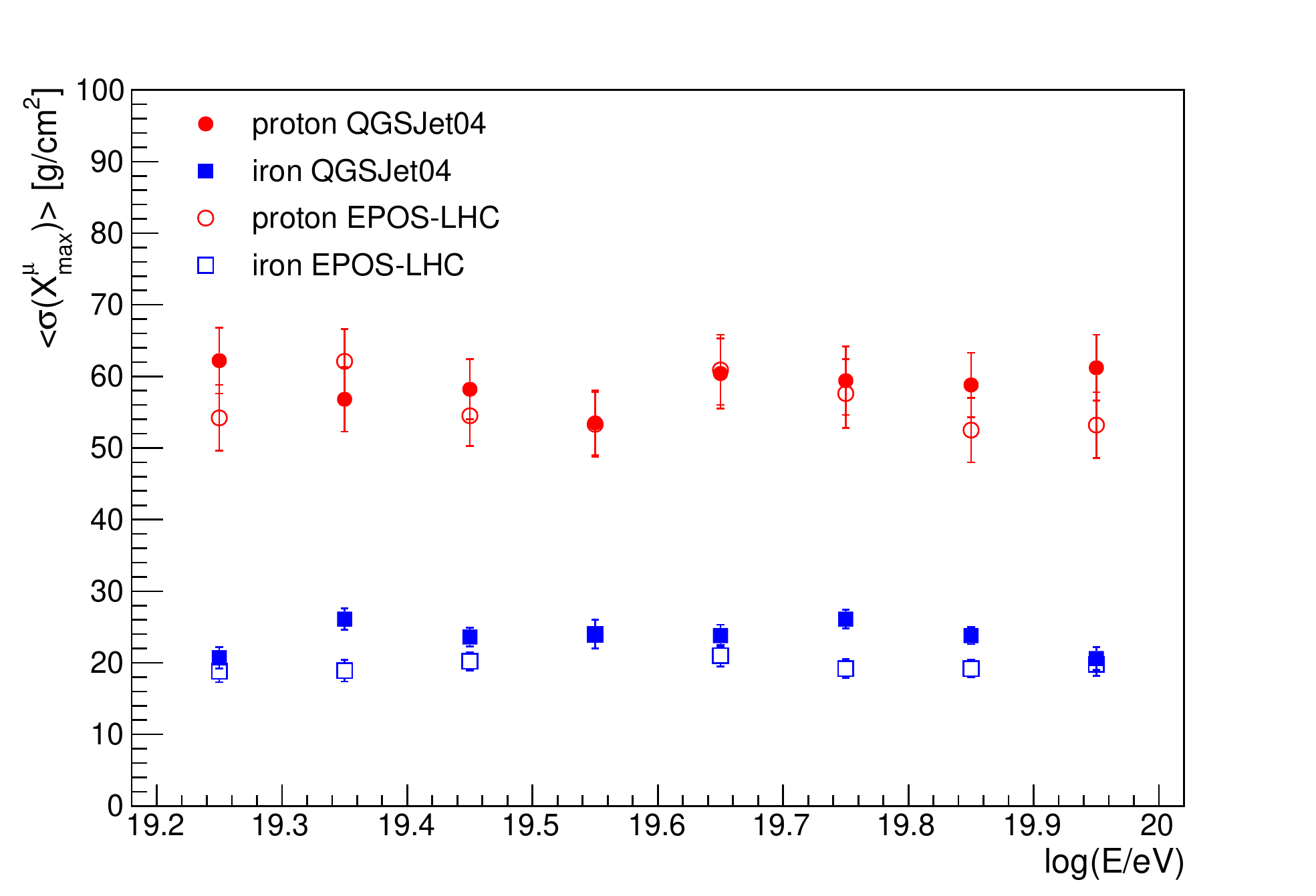}
\end{minipage}
\caption{\small Left) $\langle \sigma(X_{\mathrm{max}}^{\mu})\rangle$ is shown as a function of the zenith angle. Right) $\langle \sigma(X_{\mathrm{max}}^{\mu})\rangle$ is shown as a function of the logarithm of energy. Results are shown for proton (red) and iron (blue) showers simulated with the two post-LHC models.}
\label{sigma_zen_energy}
\end{figure}
Another important observable sensitive to primary mass is $\sigma(X_{\mathrm{max}}^{\mu})$. The standard deviation of $X_{\mathrm{max}}^{\mu}$ is smaller for heavy primaries and this property can be exploited to infer additional information about the composition of UHECRs.
The evolution of $ \sigma(X_{\mathrm{max}}^{\mu})$ with the zenith angle and the energy is shown in Fig.~\ref{sigma_zen_energy}.
Contrarily to $X_{\mathrm{max}}^{\mu}$, $\sigma(X_{\mathrm{max}}^{\mu})$ increases with the zenith angle and does not depend on the energy. 
More importantly, model predictions in the case of $\sigma(X_{\mathrm{max}}^{\mu})$ are almost in agreement between each other, which makes this observable more suitable than $X_{\mathrm{max}}^{\mu}$ for composition studies.

\section{The muon production depth reconstruction in a wide energy and angular range}\label{sec:method}

The signals recorded by SD stations are given by both muons and electromagnetic particles. To reconstruct the muon production depths for a given event, the selection of the signal solely due to muons must be performed. In this context, the electromagnetic signal is treated as a background that must be eliminated. \\
The Pierre Auger Collaboration already published a study of the muon production depths for inclined events ($\theta=55^{\circ}-65^{\circ}$) by using stations far from the core ($r>1700$ m), exploiting the fact that within these cuts the electromagnetic component is very small and can be removed with a simple threshold cut \cite{bib:MPD}.  \\
To extend the range of applicability of the MPD analysis to lower zenith angles, it is necessary to perform the analysis closer to the shower core such that enough muons are sampled.
A more refined technique to properly take into account the electromagnetic component must be used, since the electromagnetic contribution is more and more important as the zenith angle decreases, especially close to the shower core. \\
As will be described in Sect.~\ref{sec:em}, the estimation of the electromagnetic component from the total signal is achieved by means of an algorithm which exploits the different temporal structures of the electromagnetic signal and the muonic one. The method allows one to extract the time distribution of the muon signal station by station in the whole range of distances considered, i.e. $1200-4000$ m, and zenith angles, i.e. $45^{\circ}-65^{\circ}$. \\
As discussed in Sect.~\ref{sec:featMPD}, a cut at $r=1200$ m ensures a reasonably small contribution of the kinematic delay to the total one, while the angular range is chosen to have the muon maximum defined on an event-by-event basis for both proton and iron showers. \\
An energy threshold must also be set to guarantee a good determination of the $X_{\mathrm{max}}^{\mu}$ observable. In particular, we select events with  $E>10^{19.2}$ eV in order to have enough muons, i.e. $N_{\mu}>15$, to build the longitudinal profile.

\subsection{The kinematic delay parametrisation}
\label{sec:param}

To reconstruct the muon production height by means of the time model discussed in Sect.~\ref{sec:model}, the kinematic delay contribution must be evaluated muon-by-muon.
The energy carried by each single muon in the SD stations cannot be measured, and therefore the kinematic delay can be estimated only by parametrising it relying on simulations.
A modelling of the muon kinematic delay was studied to this aim in \cite{bib:mod-old}. A new parametrisation is now derived considering the new post-LHC hadronic models and the extended ranges in zenith angle and distances from the core. \\ 
To take into account the different dependences of the kinematic delay for the different masses and models, we have considered a mixed sample of primaries (50\% proton and 50\%iron) and hadronic interaction models (50\% EPOS-LHC and 50\% QGSJetII-04 for each primary mass). The parametrisation has been tuned for zenith angles between $45^{\circ}$ and $65^{\circ}$ and at a given energy, $\mathrm{log}_{10}(\mathrm{E/eV})=19.55$, since the kinematic delay does not depend significantly on the primary energy. \\
A multiparametric fit is used to derive the dependence of the kinematic delay on the distance $r$ from the core, the muon production height $z_{m}$ measured by means of the time model (see Sect.~\ref{sec:model}), and the zenith angle $\theta$. This approach allows to correctly include all the correlations among the observables. \\
Expressing the production distance as $z_{m}-\Delta$, where $\Delta$ is the distance of the muon impact point at the ground to the shower plane (see Fig.~\ref{geomdelay}(left)), the relation between the kinematic delay $t_{\epsilon}$ and the three observables is given by 
\begin{equation}
\begin{aligned}
\begin{split}
t_{\epsilon} = 53.2- 75.7  \ (z_{m}-\Delta)_{n}+ 77.4  \ z_{m}^2+ 49.7 \ r_{n} \\
-73.9 \ (z_{m}-\Delta)_{n}^{3}- 46.5  \ r_{n} \ (z_{m}-\Delta)_{n}+ 5.2 \ \theta_{n} \\
-7.2  \ (z_{m}-\Delta)_{n}  \theta_{n} + 30.8 \  (z_{m}-\Delta)_{n}^{4} \\
- 1.1 \ (z_{m}-\Delta)_{n}^{2}  \theta_{n}- 1.4 \ r_{n} \ (z_{m}-\Delta)_{n}^{2}, \\
\label{kin-par}
\end{split}
\end{aligned}
\end{equation}
where
\begin{equation}
\begin{aligned}
\begin{split}
r_{n}	= 1 + 2 \ (r - 4000) / (4000 - 1000); \\
(z_{m}-\Delta)_{n} = 1 + 2 \ ((z_{m}-\Delta)- 40000) / (40000 - 6.6); \\
\theta_{n}= 1 + 2  \ (\theta - 64.8) / (64.8 - 55.); \\
\end{split}
\end{aligned}
\end{equation}
are the normalised variables.\\
\begin{figure}
\centering
\includegraphics [width=1.\textwidth]{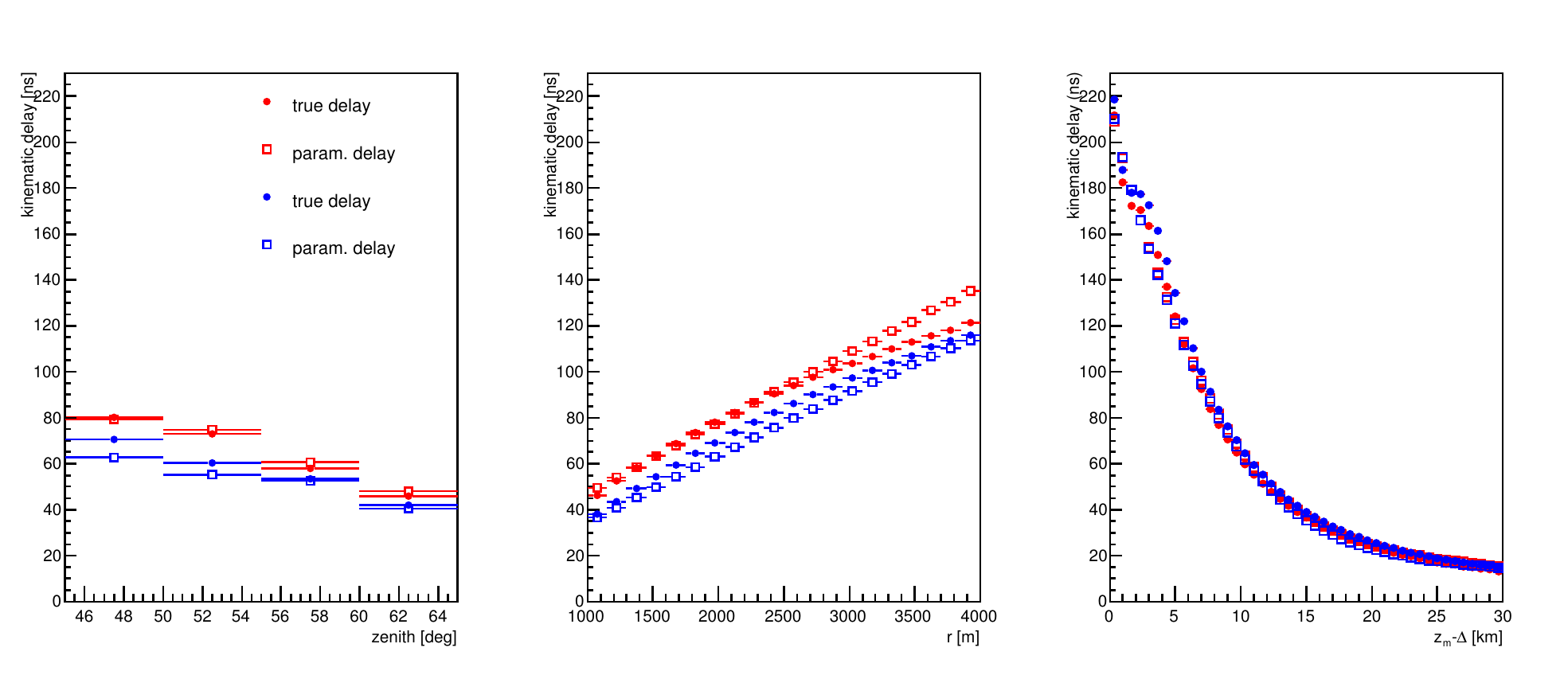}
\caption{\small The true and the parametrised kinematic delay as a function of the zenith angle (left), distance from the core (center) and production distance $z_{m}-\Delta$, in the case of proton showers simulated with EPOS-LHC model (red) and iron showers simulated with QGSJetII-04 model (blue).}
\label{kinpar1}
\end{figure}
In Fig.~\ref{kinpar1} (left) the evolution of the true and the parametrised kinematic delay is shown as a function of the zenith angle $\theta$. The kinematic delay strongly depends on $\theta$ since the energy spectrum of muons which arrive at ground level depends on the zenith angle. The parametrisation described above correctly reproduces the true kinematic delay for the two extreme cases: the model and primary giving the lowest (proton, EPOS-LHC) and the highest (iron, QGSJetII-04) shower maximum. 
Both the dependences on the distance from the core $r$ and on the production height $z_{m}-\Delta$ with respect to the shower plane are reasonably well reproduced by the parametrisation, as shown in Fig.~\ref{kinpar1} (center-right).
For distances greater than 3000 m the parametrisation performance is not optimal for proton showers. However, the kinematic delay contribution to the total one is very small at these distances, as well as the number of muons involved (less than 1\% of the total number).\\
\begin{figure}
\centering
\includegraphics [width=.8\textwidth]{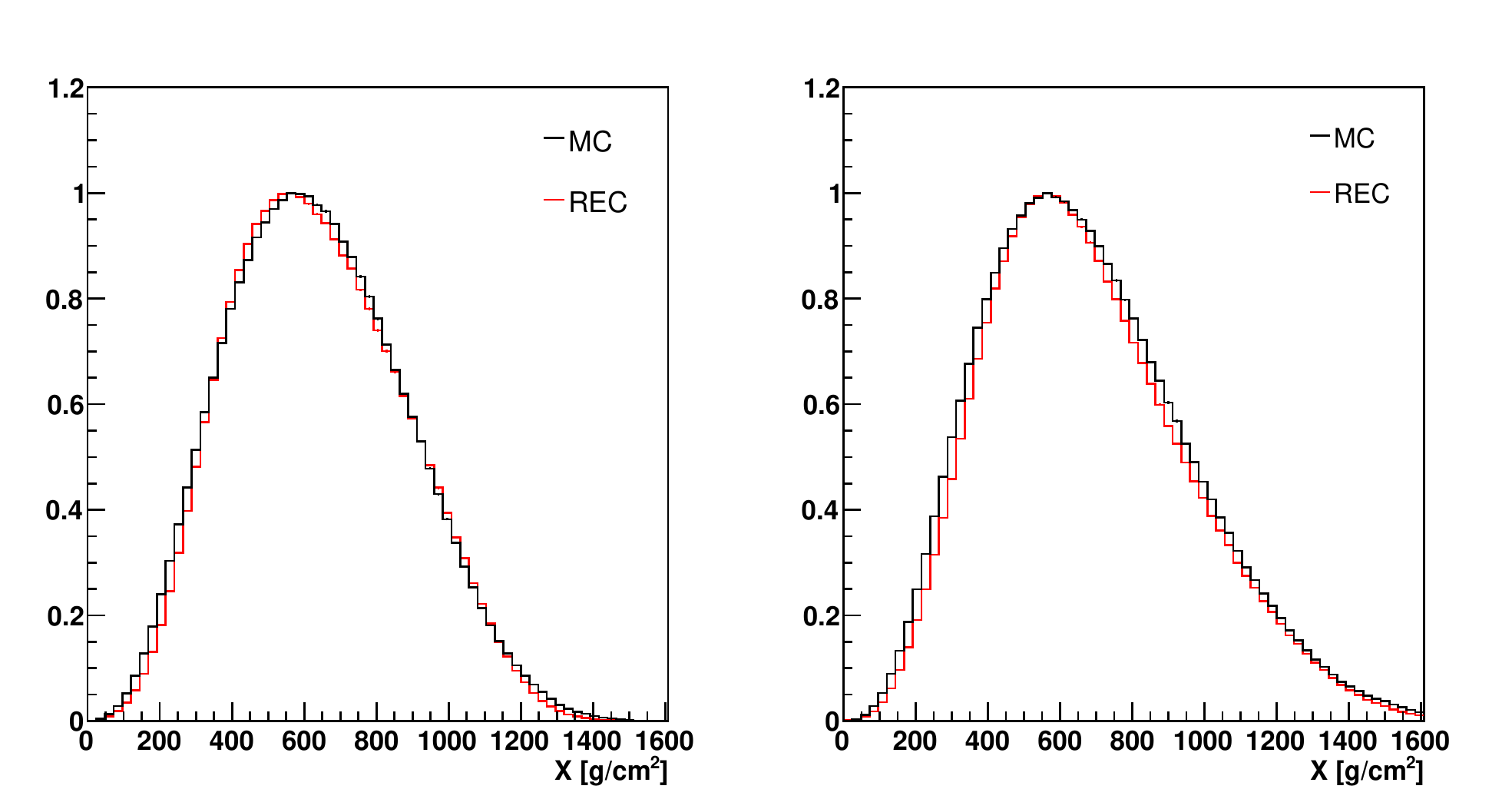}
\caption{\small  The average reconstructed MPD distribution at generation level is compared to the expected one. The results are shown for proton showers with $E=10^{19.5}$ eV in the angular $\theta=45^{\circ}-55^{\circ}$ (left) and $\theta=55^{\circ}-65^{\circ}$ (right), simulated using QGSJetII-04 model.}
\label{meanMPD}
\end{figure}
As an example, the expected and reconstructed average MPD distributions for proton showers simulated with QGSJetII-04 model at energy $10^{19.5}$ eV are shown in Fig.~\ref{meanMPD}. To point out the angular dependences of the MPD distributions we choose to show the average profiles by dividing the angular range in two bins:  $\theta=45^{\circ}-55^{\circ}$ and  $\theta=55^{\circ}-65^{\circ}$.
The reconstructed MPD distribution is obtained by applying the time model and the kinematic delay parametrisation at generation level, i.e. without detector effects. The expected and reconstructed distributions are compatible for all muon depths, in both angular ranges.

\subsection{The estimation of the electromagnetic component}
\label{sec:em}

The electromagnetic and the muonic components produce signals in the SD stations with different time distribution and this feature can be exploited to disentangle them. In particular the electromagnetic signal smoothly changes as a function of time, while the muon one is characterised by high narrow peaks. \\
The time distribution of the trace is related to the height of the shower development above the detecting surface, while the signal structure depends on the energy with which particles hit the water-Cherenkov station, on their number density, and on the light diffusion.\\
In the case of muons, the spread of arrival times at the ground is narrow since, once produced, muons no longer interact and follow more or less a straight trajectory. In addition, because of their low number density and the high energy (about 1 GeV/muon at the ground for UHECRs), muons produce peaked signals. On the contrary, electromagnetic particles are part of a cascade and the time spread at the ground is large. The high number density together with a mean energy of some 20 MeV/particle give rise to smooth signals.\\
These two kinds of signal can be separated by using a moving average algorithm, which exploits the smoothed nature of the electromagnetic trace to extract it from the total trace of each station \cite{bib:collica}.
The procedure to derive the electromagnetic and the muonic component in a time interval $T$ divided into $N_{\mathrm{bin}}$ equidistant bins is thus the following: 
\begin{itemize}
\item the total signal $S_{\mathrm{tot}}(\mathrm{i})$ in the $\mathrm{i}_{\mathrm{th}}$ is averaged over 3 well behaving PMTs;
\item the smoothed trace $S^{\mathrm{Smoo}}_{\mathrm{EM}}(t_{\mathrm{i}})$ is derived by substituting each bin content with the average value estimated in the range [$\mathrm{i}$-$N_{\mathrm{bin}}$,$\mathrm{i}$+$N_{\mathrm{bin}}$] and assigned to the electromagnetic component;
\item the muon trace of the considered bin is given by the positive difference $S^{\mathrm{Smoo}}_{\mu}(t_{\mathrm{i}}) = S_{\mathrm{tot}}(t_{\mathrm{i}})-S^{\mathrm{Smoo}}_{\mathrm{EM}}(t_{\mathrm{i}})$, if any.
\end{itemize}
$N_{\mathrm{bin}}$ is the convolute range, which depends on zenith angle and its value must be carefully tuned. If one considers the smoothing average in the frequency domain as a low-pass filter, a large convolute range will be enough to follow the small electromagnetic signal in inclined showers. On the contrary, in vertical showers a narrow window is needed.
The size of convolute range has been obtained by minimising the relative difference between the original electromagnetic signal from the simulation and the one obtained from the smoothing method.\\
The procedure is repeated $N_{\mathrm{iter}}$ times: each time, the starting signal is the original one after subtraction of the non-electromagnetic contribution obtained in the previous iteration. Few iterations are enough to reach convergence. \\
All methods which identify the muon component from the peaks they produce in the time trace have to deal with the physical background due to high-energy electromagnetic particles ($E>300$ MeV), which can produce spikes indistinguishable from those of muons.
This background is unavoidable and its impact on the muon signal reconstruction has been evaluated. In particular, the density of these particles with respect to muons ($E_{\mu}>400$ MeV) decreases with the zenith angle and the distance from the core since the electromagnetic component is more and more absorbed in the atmosphere. On the other hand, this background is more important as the energy increases. \\
In the range of applicability of the MPD reconstruction, the density of the high-energy electromagnetic particles with respect to muons is smaller than 10\%.
This background affects only the tail of the signal time distribution in the SD stations, where muons are few with respect to the high-energy electromagnetic particles. To reduce this background, a time cut is performed at the \textit{300th} bin of the FADC trace, which is approximately 1500 ns after the start time of the trace. This cut has been tuned such that more than 85\% of the muon signal is kept and thus providing enough muons to correctly reconstruct the MPD distribution. \\
The smoothing algorithm, combined with the time cut, allows to derive the muon time distribution with an accuracy of about 1\% at the beginning of the trace and of about 6\% in the last part. This is valid in the whole energy and angular range of interest, and for both masses and hadronic interaction models considered.
Regarding the extraction of the muon signal, the accuracy of the method is about 10\% for zenith angles smaller than $55^{\circ}$ while for showers with zenith angles greater than $55^{\circ}$ the method has an accuracy within 5\%. As the muon signal enters in the MPD distribution as a weight only, an accuracy within 10\% is considered satisfactory.

\subsection{The detector effects removal}
\label{sec:det}

\begin{figure}
\centering
\includegraphics [width=.8\textwidth]{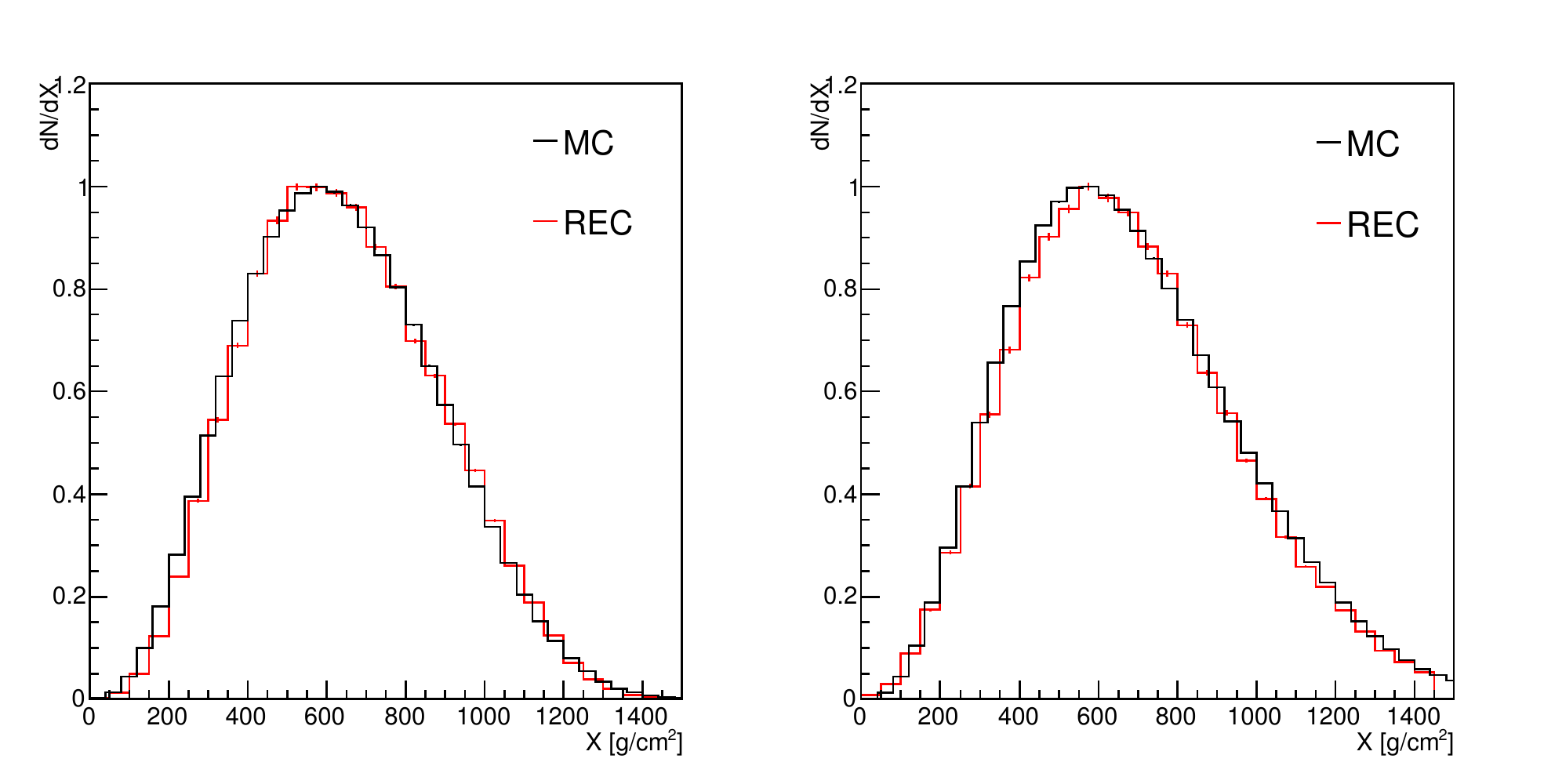}
\caption{\small The average reconstructed MPD distribution at detector level compared to the expected one. Proton showers simulated with QGSJetII-04 model at $E=10^{19.5}$ eV and for the angular range $\theta=45^{\circ}-55^{\circ}$ (left) and $\theta=55^{\circ}-65^{\circ}$ (right). }
\label{recMPD}
\end{figure}

At generation level, all muons arriving at the ground between 1200 m and 4000 m from the shower core participate to the reconstruction of the MPD distribution.
At reconstruction level the detector effects come into play, namely the discrete sampling at the ground, the time uncertainty of the SD station and its response to muons. \\
The discrete sampling affects the MPD reconstruction since it induces fluctuations due to the different muon samples. Indeed, the finite area of the detectors (10 $\mathrm{m^2}$ cross section for vertical incidence) together with their discrete grid at the ground strongly limits the number of muons being collected.
In addition, the shape of the MPD distribution observed from different positions at ground level varies because of differences in the probability of in-flight decay and because muons are not produced isotropically from the shower axis. This effect has also an impact on the reconstruction, since the estimation of the MPD distribution is obtained by integrating all muons over the distance from the core $r$. \\
The time resolution of the detector $\delta t$ gives rise to an uncertainty in the reconstruction of $X^{\mu}$ since every time bin of the temporal trace is converted into a MPD entry by means of Eqs.~\ref{eq:z-for} and \ref{eq:int}.
By assuming an exponential atmospheric density, $\rho(z) = \rho_{0}\exp(-z\cos\theta/h_{0})$, the uncertainty in $X^{\mu}$ is given by
\begin{equation}
\delta X^{\mu} = \frac{2X^{\mu}h_{0}}{r^2\cos\theta} \ln^2 \left( \frac{X^{\mu}\cos\theta}{h_{0}\rho_{0}}\right) c\delta t.
\label{eq:timeres}
\end{equation}
The impact of the time resolution on the estimation of the muon production depth is greater close to the core and for high zenith angles.\\
Lastly, the light propagation inside the detector and the electronics response smear the muon arrival times. In particular, the typical width of a muon signal is larger than a single time bin since the photoelectrons produced by muons arrive at the photomultiplier tube according to an exponential law in time, which reflects the attenuation in water and the multiple reflections off the Tyvek.
This effect causes an uncertainty in the arrival time of muons which results in an uncertainty in $X^{\mu}$.\\
The distortion in the reconstructed MPD distribution produced by these effects can be compensated by introducing a global time offset, $T_{\mathrm{shift}}=60 \ \mathrm{ns}$, to be subtracted to each time bin. Its value is constant in the whole energy and angular range and it is related to the decay time of the muon signal in the SD station.\\
In Fig.~\ref{recMPD} the average MPD distribution after the whole reconstruction chain is compared to the expected one for the angular ranges $\theta=45^{\circ}-55^{\circ}$ and $\theta=55^{\circ}-65^{\circ}$. The reconstructed and expected distributions are compatible, demonstrating that the time offset removes on average the detector effects.

\subsection{The resampling procedure}
\label{sec:resam}

To cope with the huge amount of particles in EAS, Monte Carlo codes exploit the thinning algorithm \cite{bib:hillas}, tracing to the ground only a small fraction of statistically weighted particles with the obvious benefits of a reasonable CPU time and a limited storage for the simulations. A resampling (or unthinning) algorithm is then implemented to recover the actual flux of particles \cite{bib:billoir}. 
The choice of the thinning level does not affect the MPD distribution, while the effect of the resampling turns out to be important, as reported in \cite{bib:erratum}. In particular, a non-optimal value of the resampling area in the simulations causes an underestimation of the muon delay with respect to the arrival time of the shower front.
This underestimation increases with the distance from the core and does not depend on the nature of the primary, its energy and hadronic interaction model. A time correction has thus been tuned by studying the underestimation of the muon delays in our range of distances from the core and zenith angles, with proton showers simulated with QGSJetII-04 model. Having derived the muon delay underestimation at generation level, the measured delays are corrected muon by muon at reconstruction level, where this problem appears.

\subsection{Selection criteria and performance of the method}
\label{sec:bias}

\begin{figure}
\begin{minipage}[b]{.40\textwidth}
\centering
\includegraphics [width=1.3\textwidth]{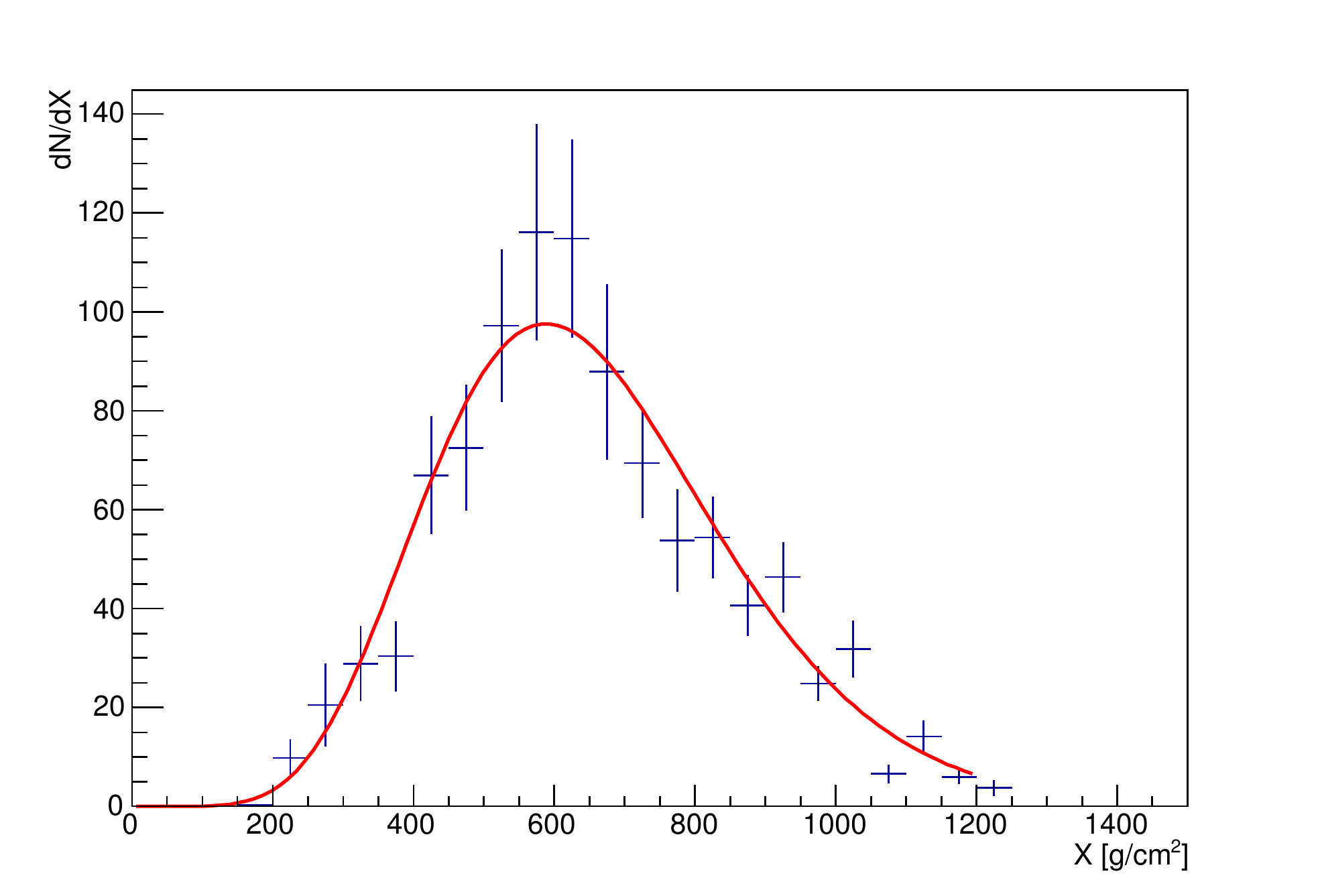} 
\end{minipage}%
\hspace{20mm}%
\begin{minipage}[b]{.40\textwidth}
\includegraphics [width=1.32\textwidth]{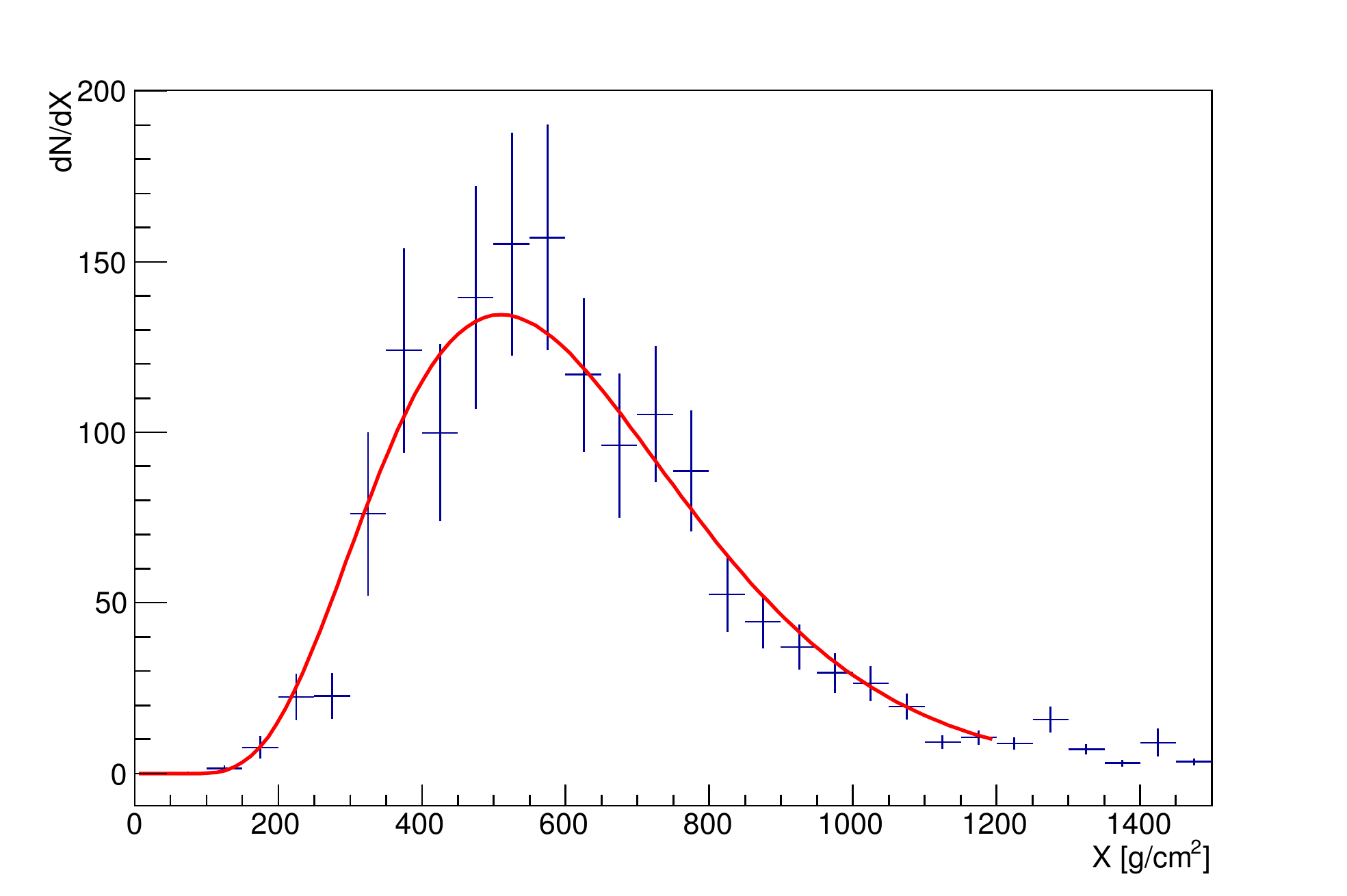}
\end{minipage}
\caption{\small Left) The reconstructed MPD distribution for a proton shower with $\theta=48^{\circ}$ and  $E=10^{19.8}$ eV simulated using QGSJetII-04 model.
Right) The reconstructed MPD distribution for a proton shower with $\theta=58^{\circ}$ and $E=10^{19.6}$ eV simulated using QGSJetII-04 model. The USP function fits are also shown.}
\label{eventsim}
\end{figure}

\begin{figure}
\begin{center}
   {\includegraphics[width=1.\textwidth]{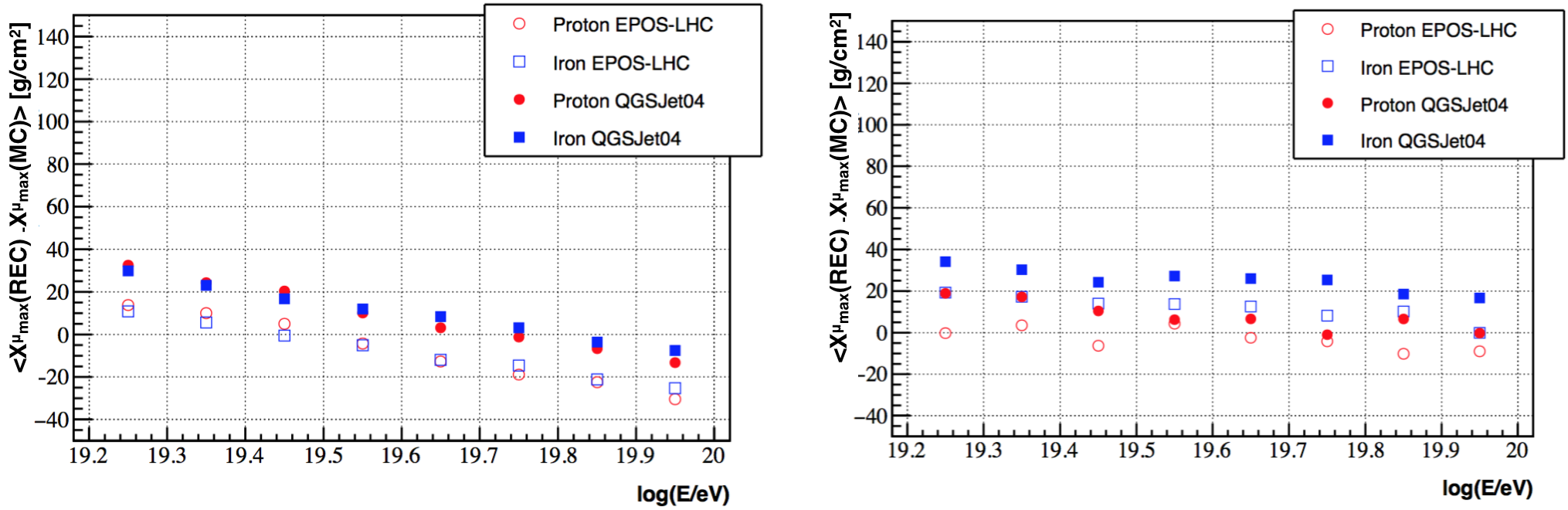}} 
\vfill
   {\includegraphics[width=1.\textwidth]{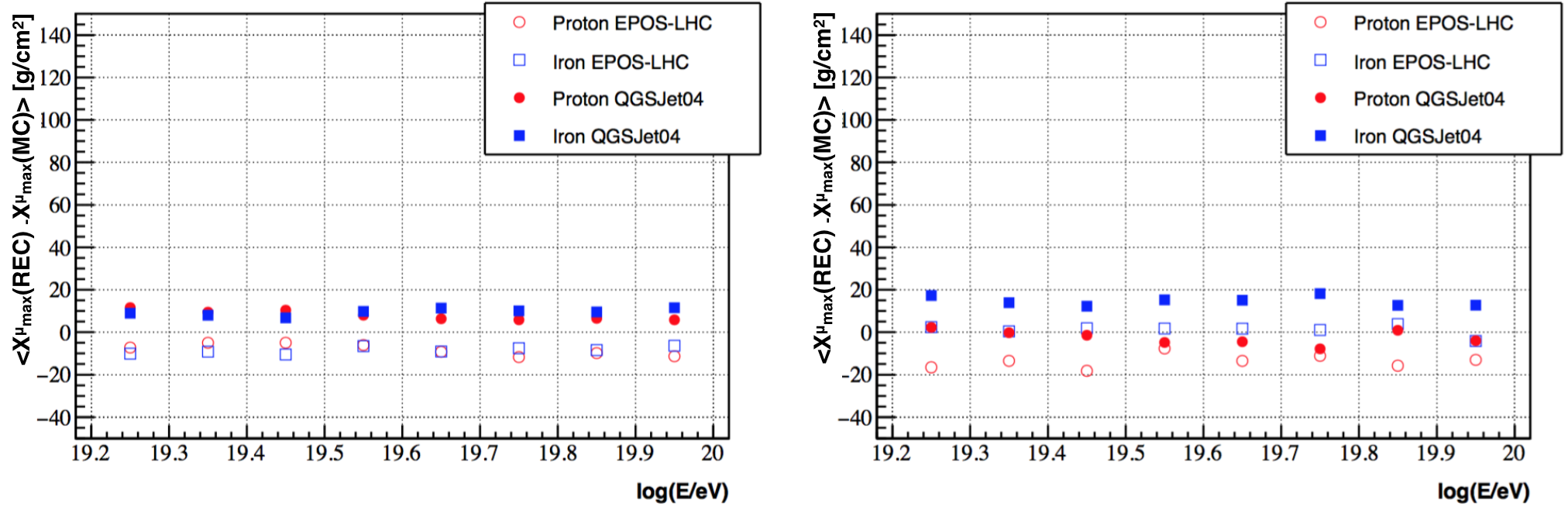}}
\caption{\small The evolution with energy of the mean of the distribution $X_{\mathrm{max}}^{\mu}(\mathrm{reconstructed})-X_{\mathrm{max}}^{\mu}(\mathrm{MC})$, before (top) and after (bottom) the correction for the angular and energy dependence of the reconstruction bias. The results are shown for the angular range $\theta=45^{\circ}-55^{\circ}$ (left) and $\theta=55^{\circ}-65^{\circ}$ (right).}\label{recbias-en}
\end{center} 
\end{figure}

\begin{figure}
\centering
\includegraphics [width=1.\textwidth]{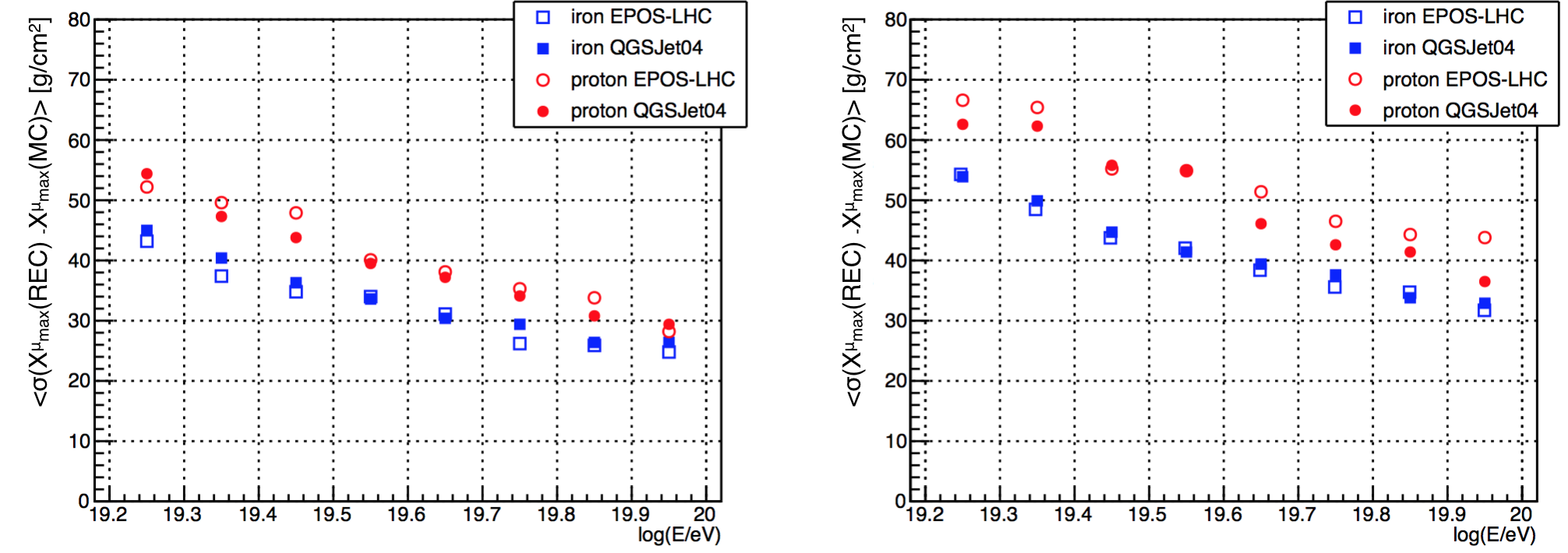}
\caption{\small The mean of $\sigma(X_{\mathrm{max}}^{\mu}(\mathrm{reconstructed})-X_{\mathrm{max}}^{\mu}(\mathrm{MC}))$ is shown as a function of energy, for the angular range $\theta=45^{\circ}-55^{\circ}$ (left) and $\theta=55^{\circ}-65^{\circ}$ (right).}
\label{det-res-mc}
\end{figure}

From the set of events with a reconstructed MPD distribution, we select those with a reliable measurement of longitudinal profile.\\
At event level, we require that the detector with the highest signal has all six closest neighbours operating.
This is a quite stringent cut to guarantee a good reconstruction of the shower parameters at the ground. Besides, we require that at least five stations with a signal greater than 3 VEM contribute to the reconstruction. This is justified to avoid trigger fluctuations and minimise the impact of accidental signals. \\
Once the fit of the MPD distribution is performed, the fit convergence is required and events with relative uncertainty $\delta X_{\mathrm{max}}^{\mu}/X_{\mathrm{max}}^{\mu}$ greater than a certain value $\epsilon_{\mathrm{max}}$ are rejected. This upper limit, quantitatively reported in table \ref{tab:error}, is an energy-dependent quantity since the accuracy in the estimation of $X_{\mathrm{max}}^{\mu}$ improves with energy. This is a natural consequence of the increase in the number of muons that enter the MPD distribution as the energy grows. The value chosen for $\epsilon_{\mathrm{max}}$ is a three-sigma limit on the $\delta X_{\mathrm{max}}^{\mu}/X_{\mathrm{max}}^{\mu}$ distribution and ensures no selection bias between the different primary species.\\
Finally, we accept events with the following condition for the parameter $L$: $130 < L < 415 \ \mathrm{g/cm^2}$. This selection cut allows to discard events for which the MPD distribution is not reconstructed well in the first part or in the tail, leading to unphysical values for the parameter $L$.\\
Two examples of a reconstructed MPD distribution are shown in Fig.~\ref{eventsim}. \\
\begin{table}[h]
\begin{center}
\begin{tabular}  {c|c}
$\mathrm{log_{10}(E/eV)}$ & $\epsilon_{\mathrm{max}}$ \\
\hline
$19.2-19.5$ & 9\%  \\
$19.5-20.$ & 6\%  \\
\hline
\end{tabular}
\end{center}
\caption{\small The maximum relative uncertainties allowed in the estimation of $X_{\mathrm{max}}^{\mu}$.} \label{tab:error}
\end{table}
The overall selection efficiency, i.e. the number of events which pass the selection cuts, is 94\% at low zenith angles and 96\% at high zenith angles. The difference in the selection efficiency for the two primaries is smaller than 5\% at all energies, zenith angles, and for both hadronic interaction models. \\ 
To evaluate the method performance, the reconstruction bias $\langle X_{\mathrm{max}}^{\mu}(\mathrm{reconstructed})-X_{\mathrm{max}}^{\mu}(\mathrm{MC}) \rangle$ has been studied and its dependence on the energy and the zenith angle has been evaluated.\\
\begin{figure}
\centering
\includegraphics [width=.6\textwidth]{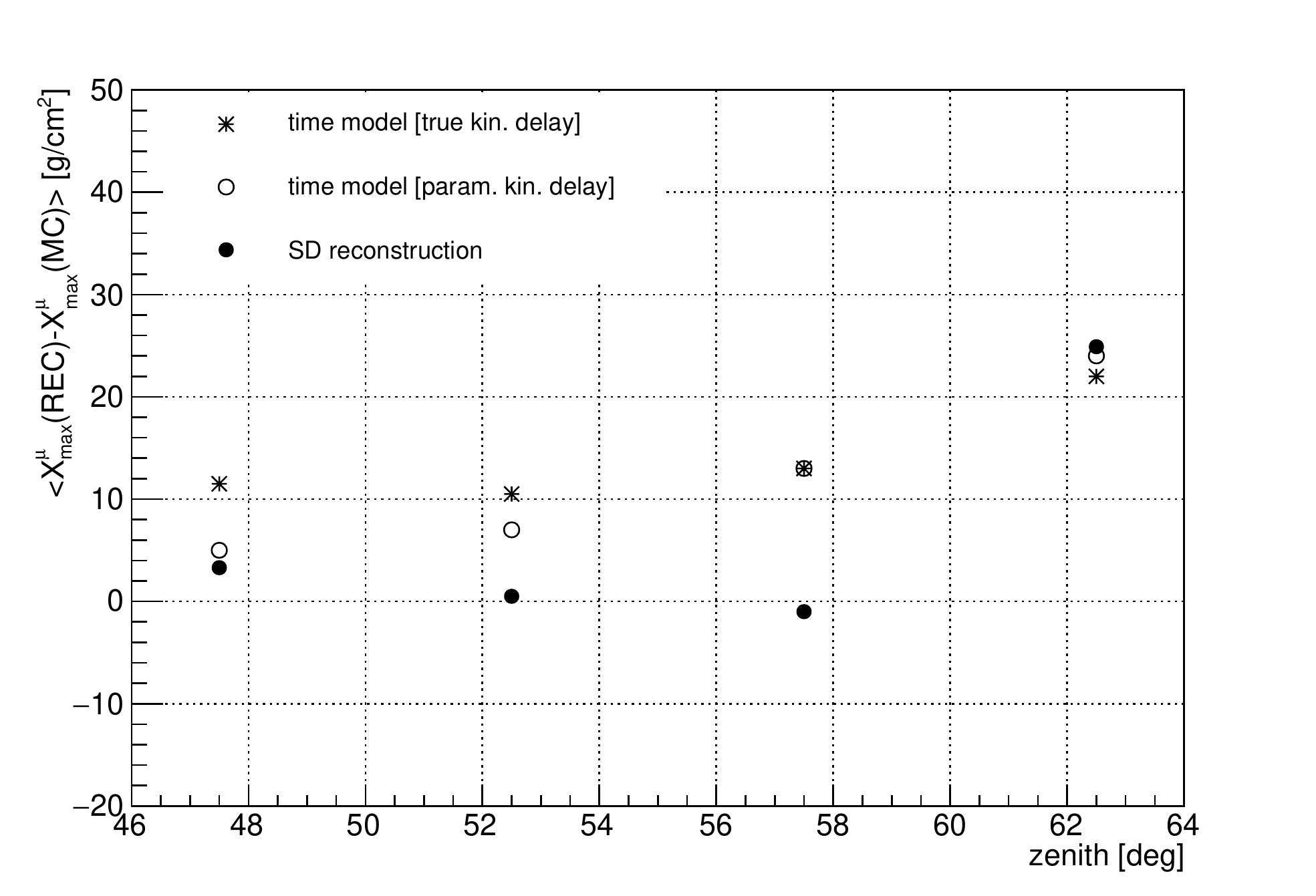}
\caption{\small The evolution with zenith angle of the mean of the distribution $X_{\mathrm{max}}^{\mu}(\mathrm{reconstructed})-X_{\mathrm{max}}^{\mu}(\mathrm{MC})$ is shown for three different cases: the time model is applied at CORSIKA level, where all the information of the muon is available, by using the true kinematic delay (stars) and the parametrised kinematic delay (empty circles); the time model and the kinematic delay parametrisation are used at detector level with the correction for the detector effects (filled circles). The results are obtained by averaging the reconstruction bias values for iron and proton showers, with both EPOS-LHC and QGSJetII-04 models.}
\label{recbias-zen}
\end{figure}
In Fig.~\ref{recbias-zen} the reconstruction bias at different steps of reconstruction is shown as a function of the zenith angle. To simplify the comparison, its angular dependence is integrated over energy, masses and hadronic interaction models. As one can see, the approximations done in the muon arrival time model introduces a bias in the reconstruction, which increases with the zenith angle. Once that the reconstruction with the SD is performed, and the time offset is applied, the reconstruction bias becomes quite compatible with the one at generation level, ranging from about 4 $\mathrm{g/cm^2}$ to 24 $\mathrm{g/cm^2}$ as the zenith angle increases. The reconstruction bias is smaller at detector level than at generation level for zenith angles up to $60^{\circ}$ because a unique time offset is used despite the large angular range.
However, the chosen value allows to fairly reproduce the angular dependence of the bias at generation level.\\
The summary of the values for the reconstruction bias in the different analysis steps is reported in table \ref{tab:bias}, averaged for the two zenith range $\theta=45^{\circ}-55^{\circ}$ and $\theta=55^{\circ}-65^{\circ}$.\\
\begin{table}[h]
\begin{center}
\begin{tabular}  {c|c|c}
 & \multicolumn{2}{c}{{Rec. bias [$\mathrm{g/cm^{2}}$]}}\\
Analysis step &   $45^{\circ}-55^{\circ}$ & $55^{\circ}-65^{\circ}$ \\
\hline
time model [true kin. delay] &  11 & 17  \\
time model [param. kin. delay] &  5 & 15\\
SD reconstruction &  2 & 12  \\
\hline
\end{tabular}
\end{center}
\caption{\small The average reconstruction bias in the different analysis steps for the two angular ranges.}
 \label{tab:bias}
\end{table}
In Fig.~\ref{recbias-en} (top) the reconstruction bias is shown as a function of the energy, for different primaries and hadronic interaction models. As before, the results are shown for the two zenith range $\theta=45^{\circ}-55^{\circ}$ and $\theta=55^{\circ}-65^{\circ}$.
The bias decreases with the energy, especially at low zenith angles, and this effect is related to the extraction of the muon signal in the SD stations. Indeed, as the energy increases more muons enter the stations producing overlapped signals which causes an energy-dependent performance of the method.\\
Both the angular and the energy dependences are taken into account by correcting the measured $X^{\mu}_{\mathrm{max}}$ with tabulated values of the reconstruction bias, estimated for the different energies and zenith angles. The final bias evaluated after the whole reconstruction chain and after the correction for the energy and angular dependence is shown in Fig~\ref{recbias-en} (bottom).\\
\begin{table}
\begin{center}
\begin{tabular}  {c|c|c}
 & \multicolumn{2}{c}{{Mass and model spread [$\mathrm{g/cm^{2}}$]}}\\
 Analysis step &  $45^{\circ}-55^{\circ}$ &  $55^{\circ}-65^{\circ}$ \\
\hline
time model [true kin. delay] & $<$1  & $<$1 \\
time model [param. kin. delay] & 5  & 10\\
SD reconstruction & 10  & 15  \\
\hline
\end{tabular}
\end{center}
\caption{\small The mass and model spread in the different analysis steps for the two angular ranges.} \label{tab:spread}
\end{table}
The method shows a different performance depending on the primary and the hadronic interaction model. In particular the reconstruction bias is positive in the case of iron showers simulated with the QGSJetII-04 model and negative in the case of proton showers simulated with the EPOS-LHC model, staying within 15 $\mathrm{g/cm^{2}}$ ($\theta=45^{\circ}-55^{\circ}$) and 20 $\mathrm{g/cm^{2}}$ ($\theta=55^{\circ}-65^{\circ}$), as shown in Fig.~\ref{recbias-en} (bottom).
The mean difference between the maximum and the minimum bias values is defined as the \textit{mass and model spread} and must be taken into account as a systematic uncertainty. \\
In table ~\ref{tab:spread} the spread for the different analysis steps is reported.
The mass and model spread arises when the kinematic delay parametrisation is introduced, since the parametrisation is tuned on a mixed sample.
When the SD reconstruction is performed, the spread increases in both angular range. As Fig.~\ref{recbias-en} (bottom) shows, the spread is mainly due to hadronic interaction models up to $55^{\circ}$, amounting to about $10 \ \mathrm{g/cm^{2}}$.  For higher angles, it increases to 15 $\mathrm{g/cm^{2}}$ due to the non-negligible contribution of the mass spread. \\
To reduce the spread, two options could be followed.
Firstly, the parametrisation could be tuned on a single mass and model case, for example on proton showers simulated with QGSJetII-04 model. In that case, the spread would be few $\mathrm{g/cm^{2}}$ smaller, at the cost of an additional contribution to the systematic uncertainty.
Secondly, the kinematic delay contribution to the total muon delay could be reduced by moving the distance cut to $r_{\mathrm{cut}} {>} 1500$ m. Following this approach, the spread would be few $\mathrm{g/cm^{2}}$ smaller but with a significant loss of events (about 50\%) in the lower energy range $10^{19.2}-10^{19.4}$ eV.
Since both options do not reduce significantly the spread and introduce some drawbacks, we decided not to exploit them.\\
The quality of the MPD profile reconstruction improves with the number of sampled muons. As a consequence, the performance of the method depends on the primary energy, the zenith angle and the nature of the primary particle.\\
The resolution of the method, which is given by the standard deviation of the distribution $\langle X_{\mathrm{max}}^{\mu}(\mathrm{reconstructed})-X_{\mathrm{max}}^{\mu}(\mathrm{MC}) \rangle$, is shown in  Fig.~\ref{det-res-mc} as a function of the primary energy, for both proton and iron showers simulated with QGSJetII-04 and EPOS-LHC hadronic interaction models.
The resolution improves with the energy and as the zenith angle decreases, ranging from about 50 (60) $\mathrm{g/cm^{2}}$ at $\mathrm{log_{10}(E/eV)}$=19.25 to about 25 (40) $\mathrm{g/cm^{2}}$  at $\mathrm{log_{10}(E/eV)}=19.95$ for the angular range $\theta=45^{\circ}-55^{\circ}$ ($\theta=55^{\circ}-65^{\circ}$). It is smaller in the case of iron showers, which are richer in muons. \\
The main factor which limits the resolution on $X_{\mathrm{max}}^{\mu}$ measurement is the number of muons involved in the reconstruction, which accounts for about 50\% of the total resolution. Indeed, the discrete sampling of muons at the ground worsens the resolution, especially for low energies where the number of participating stations is small.
The time uncertainty of the detector plays an important role too. The uncertainty is greater at high zenith angles and decreases rapidly with the distance from the core (see Eq.~\ref{eq:timeres}). In particular, its contribution to the resolution is about 15\% at $50^{\circ}$  and 30\% at $60^{\circ}$. 
Reconstructing the MPD profile near the core increases the contribution due to the time resolution but on the contrary decreases the more important contribution due to the number of muons sampled at the ground. Finally, the uncertainty in the determination of the core position and the angular reconstruction contributes to the total resolution at the level of about 15\%.
Of negligible importance (below 1\%) is the contribution due to the method itself, namely, the kinematic delay parametrisation. \\

The overall systematic uncertainty in the $X_{\mathrm{max}}^{\mu}$ estimation is at most 17 $\mathrm{g/cm^2}$ for the whole considered angular range \cite{bib:collica2}. This represents at most 25\% of the proton-iron separation. The most relevant contributions come from reconstruction, differences in the hadronic interaction models, and unknown primary mass.

\section{Summary} \label{sec:sum}

The Pierre Auger Observatory employs water-Cherenkov detectors to measure particle densities at the ground and therefore has a good sensitivity to the muon content of air showers. \\
By means of a model which relates the arrival time of muons at the ground with their production depths, a new approach to study the longitudinal development of the hadronic component of EAS has been established. Studying the muon profiles helps to improve our understanding of hadronic interactions at the highest energies and sets additional constraints on model descriptions. \\
In this paper we presented a method to reconstruct the muon production depth distribution on an event-by-event basis in the angular range $\theta=45^{\circ}-65^{\circ}$ and for energies $\mathrm{E}=10^{19.2}-10^{20}$ eV.\\
The large range of applicability of this analysis has been obtained thanks to the capability of extracting the muon time distribution in the SD stations for a wide range of distances from the core.
The maximum of the muon production depth distribution, $X_{\mathrm{max}}^{\mu}$, is estimated with a systematic uncertainty of at most 17 $\mathrm{g/cm^{2}}$ and its measurement could be exploited to constrain the most recent LHC-tuned hadronic interaction models, QGSJETII-04 and EPOS-LHC, and could give insights about the nature of UHECRs. In addition, thanks to the high statistics made available, deeper insights in the hadronic interaction models can be achieved by the study of the angular dependence of $X_{\mathrm{max}}^{\mu}$, and its correlation with the electromagnetic counterpart, $X_{\mathrm{max}}$.
 
 \section*{Acknowledgments}

\begin{sloppypar}
The successful installation, commissioning, and operation of the Pierre Auger Observatory would not have been possible without the strong commitment and effort from the technical and administrative staff in Malarg\"ue. We are very grateful to the following agencies and organizations for financial support:
\end{sloppypar}

\begin{sloppypar}
Comisi\'on Nacional de Energ\'\i{}a At\'omica, Agencia Nacional de Promoci\'on Cient\'\i{}fica y Tecnol\'ogica (ANPCyT), Consejo Nacional de Investigaciones Cient\'\i{}ficas y T\'ecnicas (CONICET), Gobierno de la Provincia de Mendoza, Municipalidad de Malarg\"ue, NDM Holdings and Valle Las Le\~nas, in gratitude for their continuing cooperation over land access, Argentina; the Australian Research Council; Conselho Nacional de Desenvolvimento Cient\'\i{}fico e Tecnol\'ogico (CNPq), Financiadora de Estudos e Projetos (FINEP), Funda\c{c}\~ao de Amparo \`a Pesquisa do Estado de Rio de Janeiro (FAPERJ), S\~ao Paulo Research Foundation (FAPESP) Grants No.\ 2010/07359-6 and No.\ 1999/05404-3, Minist\'erio de Ci\^encia e Tecnologia (MCT), Brazil; Grant No.\ MSMT CR LG15014, LO1305 and LM2015038 and the Czech Science Foundation Grant No.\ 14-17501S, Czech Republic; Centre de Calcul IN2P3/CNRS, Centre National de la Recherche Scientifique (CNRS), Conseil R\'egional Ile-de-France, D\'epartement Physique Nucl\'eaire et Corpusculaire (PNC-IN2P3/CNRS), D\'epartement Sciences de l'Univers (SDU-INSU/CNRS), Institut Lagrange de Paris (ILP) Grant No.\ LABEX ANR-10-LABX-63, within the Investissements d'Avenir Programme Grant No.\ ANR-11-IDEX-0004-02, France; Bundesministerium f\"ur Bildung und Forschung (BMBF), Deutsche Forschungsgemeinschaft (DFG), Finanzministerium Baden-W\"urttemberg, Helmholtz Alliance for Astroparticle Physics (HAP), Helmholtz-Gemeinschaft Deutscher Forschungszentren (HGF), Ministerium f\"ur Wissenschaft und Forschung, Nordrhein Westfalen, Ministerium f\"ur Wissenschaft, Forschung und Kunst, Baden-W\"urttemberg, Germany; Istituto Nazionale di Fisica Nucleare (INFN),Istituto Nazionale di Astrofisica (INAF), Ministero dell'Istruzione, dell'Universit\'a e della Ricerca (MIUR), Gran Sasso Center for Astroparticle Physics (CFA), CETEMPS Center of Excellence, Ministero degli Affari Esteri (MAE), Italy; Consejo Nacional de Ciencia y Tecnolog\'\i{}a (CONACYT) No.\ 167733, Mexico; Universidad Nacional Aut\'onoma de M\'exico (UNAM), PAPIIT DGAPA-UNAM, Mexico; Ministerie van Onderwijs, Cultuur en Wetenschap, Nederlandse Organisatie voor Wetenschappelijk Onderzoek (NWO), Stichting voor Fundamenteel Onderzoek der Materie (FOM), Netherlands; National Centre for Research and Development, Grants No.\ ERA-NET-ASPERA/01/11 and No.\ ERA-NET-ASPERA/02/11, National Science Centre, Grants No.\ 2013/08/M/ST9/00322, No.\ 2013/08/M/ST9/00728 and No.\ HARMONIA 5 -- 2013/10/M/ST9/00062, Poland; Portuguese national funds and FEDER funds within Programa Operacional Factores de Competitividade through Funda\c{c}\~ao para a Ci\^encia e a Tecnologia (COMPETE), Portugal; Romanian Authority for Scientific Research ANCS, CNDI-UEFISCDI partnership projects Grants No.\ 20/2012 and No.194/2012 and PN 16 42 01 02; Slovenian Research Agency, Slovenia; Comunidad de Madrid, FEDER funds, Ministerio de Educaci\'on y Ciencia, Xunta de Galicia, European Community 7th Framework Program, Grant No.\ FP7-PEOPLE-2012-IEF-328826, Spain; Science and Technology Facilities Council, United Kingdom; Department of Energy, Contracts No.\ DE-AC02-07CH11359, No.\ DE-FR02-04ER41300, No.\ DE-FG02-99ER41107 and No.\ DE-SC0011689, National Science Foundation, Grant No.\ 0450696, The Grainger Foundation, USA; NAFOSTED, Vietnam; Marie Curie-IRSES/EPLANET, European Particle Physics Latin American Network, European Union 7th Framework Program, Grant No.\ PIRSES-2009-GA-246806; and UNESCO.
\end{sloppypar}
 


\end{document}